\documentclass[aps,prd,longbibliography,reprint,twocolumn,amsmath,amssymb,amsfonts,showpacs,footnote,superscriptaddress]{revtex4-1}%reprint

\usepackage{ifpdf}
\usepackage{dcolumn}
\usepackage{enumitem}
\usepackage{bm}
\usepackage{here}
\usepackage{float}
\floatstyle{plaintop}
\restylefloat{table}
\ifpdf

      \usepackage[pdftex]{graphicx}  % note the x at the end
      \usepackage[pdftex]{epsfig}

      \usepackage[pdftex]{hyperref}
\hypersetup
	{
		colorlinks,%
		citecolor=blue,%
		linkcolor=red,%
		urlcolor=blue,%
	}

\else

      \usepackage[dvips]{graphicx}  % note the x at the end
      \usepackage[dvips]{epsfig}

      \usepackage[dvipdfm]{hyperref}
      \hypersetup
	{
		colorlinks,%
		citecolor=blue,%
		linkcolor=red,%
		urlcolor=blue,%
	}
\fi
\usepackage{gensymb} %pour le symbole Degrée

\begin{document}

\title{Regge pole description of scattering of gravitational waves \\
 by a Schwarzschild black hole}

\author{Antoine Folacci}
\email{folacci@univ-corse.fr}
\affiliation{Equipe Physique
Th\'eorique, SPE, UMR 6134 du CNRS
et de l'Universit\'e de Corse,\\
Universit\'e de Corse, Facult\'e des Sciences, BP 52, F-20250 Corte,
France}

\author{Mohamed \surname{Ould~El~Hadj}}
\email{med.ouldelhadj@gmail.com}

\affiliation{Equipe Physique
Th\'eorique, SPE, UMR 6134 du CNRS
et de l'Universit\'e de Corse,\\
Universit\'e de Corse, Facult\'e des Sciences, BP 52, F-20250 Corte,
France}

\affiliation{Consortium for Fundamental Physics, School of Mathematics and Statistics, University of Sheffield,\\ Hicks Building, Hounsfield Road, Sheffield S3 7RH, United Kingdom}

\date{\today}

\begin{abstract}

We revisit the problem of plane monochromatic gravitational waves impinging upon a Schwarzschild black hole using complex angular momentum techniques. By extending our previous study concerning scalar and electromagnetic waves [A. Folacci and M. Ould El Hadj, Phys.\, Rev.\, D {\bf 99}, 104079, (2019), arXiv:1901.03965], we provide complex angular momentum representations and Regge pole approximations of the helicity-preserving and helicity-reversing scattering amplitudes and of the total differential scattering cross section. We show, in particular, that for high frequencies (i.e., in the short-wavelength regime), a small number of Regge poles permits us to describe numerically with very good agreement the black hole glory and the orbiting oscillations and we then provide a semiclassical approximation that unifies these two phenomena.

\end{abstract}

\maketitle

\tableofcontents

\section{Introduction}
\label{Intro}

In this article, we extend to gravitational waves our previous study concerning the Regge pole description of scattering of scalar and electromagnetic waves by a Schwarzschild black hole (BH) \cite{Folacci:2019cmc}. Scattering of gravitational waves by this BH is usually tackled from partial wave methods (see Refs.~\cite{Matzner:1977dn,MatznerRyan1978,Handler:1980un,Dolan:2007ut,Dolan:2008kf} as well as the monograph of Futterman, Handler and Matzner \cite{Futterman:1988ni}) which permit us to construct numerically the helicity-preserving and helicity-reversing scattering amplitudes as well as the total differential scattering cross section. By revisiting this problem from complex angular momentum (CAM) techniques, we shall show, in particular, that they provide a powerful tool of resummation of the partial wave expansions which is helpful to describe numerically and semiclassically, in terms of Regge poles, the BH glory and the orbiting oscillations. Our approach and our results could have interesting applications in the context of strong gravitational lensing of gravitational waves. Indeed, with the recent detection of gravitational waves generated by coalescing binaries \cite{Abbott:2016blz} and with the planned development of ground-based and space-based interferometers of considerable sensitivity, it should be possible in a more or less distant future to observe gravitational signals gravitationally lensed by BHs.

It should be noted that we shall not go back on what makes CAM techniques and Regge pole analyses in BH physics interesting. We refer the reader to the introduction of our previous paper \cite{Folacci:2019cmc} and to references therein (see, in particular, Refs.~\cite{Andersson:1994rk,Andersson:1994rm,Decanini:2002ha,Decanini:2010fz,Decanini:2011xi}) as well as to the introduction of Ref.~\cite{Folacci:2018sef} where we used the CAM approach to describe the gravitational radiation generated by a particle falling radially into the Schwarzschild BH.

Our paper is organized as follows. In Sec.~\ref{SecII}, by means of the Sommerfeld-Watson transform \cite{Watson18,Sommerfeld49,Newton:1982qc} and Cauchy's residue theorem, we construct for plane monochromatic gravitational waves impinging upon a Schwarzschild BH, exact CAM representations of the helicity-preserving and helicity-reversing scattering amplitudes and of the total differential scattering cross section. These CAM representations are split into a background integral in the CAM plane and a sum over the Regge poles of the $S$-matrices corresponding to the even and odd perturbations of the BH which, in addition, involves the associated residues. In Sec.~\ref{SecIII}, we compute numerically, for various reduced frequencies, the Regge poles of the $S$-matrices, the associated residues and the background integrals. This permits us to reconstruct the scattering amplitudes and the cross section of the BH and to show that, in the short-wavelength regime, they can be described from the Regge pole sums alone with very good agreement. We also discuss the role of the background integral for low reduced frequencies, i.e., in the long-wavelength regime. In Sec.~\ref{SecIV}, from asymptotic expressions for the lowest Regge poles and the associated residues based on the correspondence Regge poles/``surface waves'' propagating close to the photon sphere \cite{Andersson:1994rm,Decanini:2002ha,Decanini:2009mu,Dolan:2009nk,Decanini:2010fz}, we provide an analytical approximation describing with very good agreement both the BH glory and a large part of the orbiting oscillations. In the Conclusion, we briefly consider possible extensions of our work.

Throughout this article, we adopt units such that $G = c = 1$. We furthermore consider that the exterior of the Schwarzschild BH is defined by the line element $ds^2= -f(r) dt^2+ f(r)^{-1}dr^2+ r^2 d\theta^2 + r^2 \sin^2\theta d\varphi^2$ where $f(r)=1-2M/r$ and $M$ is the mass of the BH while $t \in ]-\infty, +\infty[$, $r \in ]2M,+\infty[$, $\theta \in [0,\pi]$ and $\varphi \in [0,2\pi]$ are the usual Schwarzschild coordinates. We finally assume a time dependence $\exp(-i\omega t)$ for gravitational waves.

\section{Scattering amplitudes and scattering cross section for gravitational waves, their CAM representations and Regge pole approximations}
\label{SecII}

In this section, we first recall, for plane monochromatic gravitational waves impinging upon a Schwarzschild BH, the partial wave expansions of the differential scattering cross section and of the helicity-preserving and helicity-reversing scattering amplitudes. We then provide exact CAM representations of these scattering amplitudes (and therefore of the cross section) by means of the Sommerfeld-Watson transform \cite{Watson18,Sommerfeld49,Newton:1982qc} and Cauchy's theorem. These representations are split into a background integral in the CAM plane and a sum over the Regge poles of the $S$-matrices corresponding to the even and odd perturbations which, in addition, involves the associated residues.

\subsection{Partial wave expansion of the differential scattering cross section}
\label{SecIIa}

Gravitational waves impinging upon a Schwarzschild BH can be considered as gravitational perturbations of the BH, a topic which has been the subject of lot of works since the pioneering articles by Regge and Wheeler \cite{Regge:1957td} and Zerilli \cite{Zerilli:1971wd} (see also the monograph of Chandrasekhar \cite{Chandrasekhar:1985kt}). It should be recalled that these perturbations are divided into even (polar) and odd (axial) perturbations according to their even or odd parity in the antipodal transformation on the unit $2$-sphere $S^2$. Here, and in the following, we shall associate the parity symbols $p=e$ with the even perturbations and $p=o$ with the odd ones.

The differential scattering cross section for plane monochromatic gravitational waves impinging upon a Schwarzschild BH can be written in the form \cite{Dolan:2007ut,Dolan:2008kf} (see also Refs.~\cite{Matzner:1977dn,MatznerRyan1978,Handler:1980un})
\begin{equation}\label{GW_Scattering_diff}
  \frac{d\sigma}{d\Omega} = |f^{+}(\omega,x)|^2+|f^{-}(\omega,x)|^2
\end{equation}
where $f^{+}(\omega,x)$ and $f^{-}(\omega,x)$ are scattering amplitudes which are given by
\begin{equation}\label{GW_Scattering_amp}
  f^{\pm}(\omega,x) = {\widehat {\cal L}}^{\pm}_{x} \, {\widetilde f}^{\pm}(\omega,x)
\end{equation}
with
\begin{eqnarray}\label{GW_Scattering_amp_tilde}
 & &{\widetilde f}^{\pm}(\omega,x) = \frac{1}{2 i \omega} \sum_{\ell = 2}^{\infty} \frac{(2\ell+1)}{(\ell-1)\ell (\ell+1)(\ell+2)}\nonumber
 \\
 & & \times \left[ \frac{1}{2} \left(S^{(e)}_{\ell}(\omega) \pm S^{(o)}_{\ell}(\omega) \right) - \left( \frac{1 \pm 1}{2} \right) \right] P_{\ell}(x).
\end{eqnarray}
In the previous expressions, the variable $x$ is linked to the scattering angle $\theta$ by $x= \cos \theta$, the functions $P_{\ell}$ are the Legendre polynomials \cite{AS65} while the differential operators ${\widehat {\cal L}}^{\pm}_{x}$ which act on the partial-wave series (\ref{GW_Scattering_amp_tilde}) can be defined by
\begin{equation}\label{GW_Scattering_OpDiff}
{\widehat {\cal L}}^{\pm}_{x} = (1 \pm x)^2 \frac{d}{dx} \left\lbrace (1 \mp x)\frac{d^2}{dx^2} \left[ (1 \mp x)  \frac{d}{dx} \right] \right\rbrace.
\end{equation}
It should be noted that our notations for the scattering amplitudes (\ref{GW_Scattering_amp}) and (\ref{GW_Scattering_amp_tilde}) differ slightly from those that can be found in the literature. In fact, we have written these amplitudes in a form much more tractable in order to extract exact CAM representations from their partial wave expansions. It is important to recall that $f^{+}(\omega,x)$ and $f^{-}(\omega,x)$ are respectively the helicity-preserving and helicity-reversing scattering amplitudes.

We also recall that the $S$-matrix elements $S^{(p)}_{\ell}(\omega)$ appearing in Eq.~(\ref{GW_Scattering_amp_tilde}) can be defined from the modes $\phi_{\omega\ell}^{\mathrm {in} \, {(p)}}$. These modes are solutions of the homogeneous Zerilli-Moncrief (for $p=e$) and Regge-Wheeler (for $p=o$) equations
\begin{equation}
\label{H_ZMetRW_equation}
\left[\frac{d^{2}}{dr_{\ast}^{2}}+\omega^{2}-V^{(p)}_{\ell}(r)\right]\phi^{(p)}_{\omega\ell}= 0
\end{equation}
with the Zerilli-Moncrief potential given by
\begin{eqnarray} \label{pot Zerilli}
& & V_\ell^{(e)}(r)=f(r)\nonumber \\
& & \qquad \times \left[\frac{\Lambda^2(\Lambda+2) r^3+6\Lambda^2 Mr^2+36\Lambda M^2r+72M^3}{(\Lambda r+6M)^2r^3} \right]\nonumber \\
& &
\end{eqnarray}
and the Regge-Wheeler potential given by
\begin{equation} \label{pot Regge-Wheeler}
V_\ell^{(o)}(r)=f(r)\left(\frac{\Lambda+2}{r^2}-\frac{6M}{r^3} \right).
\end{equation}
In Eqs.~(\ref{pot Zerilli}) and (\ref{pot Regge-Wheeler}), we have introduced the parameter $\Lambda=(\ell -1) (\ell+2) = \ell (\ell+1)-2$. The functions $\phi_{\omega\ell}^{\mathrm {in} \, {(p)}}$ are defined by their purely ingoing behavior at the event horizon $r=2M$ (i.e., for $r_\ast \to -\infty$)
\begin{subequations}
\label{bc_in}
\begin{equation}\label{bc_1_in}
\phi^{\mathrm {in} \, {(p)}}_{\omega \ell} (r)\scriptstyle{\underset{r_\ast \to -\infty}{\sim}} \displaystyle{e^{-i\omega r_\ast}}
\end{equation}
while, at spatial infinity $r \to +\infty$ (i.e., for $r_\ast \to +\infty$), they have an
asymptotic behavior of the form
\begin{equation}\label{bc_2_in}
\phi^{\mathrm {in} \, {(p)}}_{\omega  \ell}(r) \scriptstyle{\underset{r_\ast \to +\infty}{\sim}}
\displaystyle{A^{(-,{p})}_\ell (\omega) e^{-i\omega r_\ast} + A^{(+,{p})}_\ell (\omega) e^{+i\omega r_\ast}}.
\end{equation}
\end{subequations}
In the previous expressions, the coefficients $A^{(-,{p})}_\ell (\omega)$ and $A^{(+,{p})}_\ell (\omega)$ are complex amplitudes. They are related to the $S$-matrix elements appearing in Eq.~(\ref{GW_Scattering_amp_tilde}) by
\begin{equation}\label{Matrix_S}
  S^{(p)}_{\ell}(\omega) =  e^{i(\ell+1)\pi} \, \frac{A^{(+,{p})}_\ell (\omega)}{A^{(-,{p})}_\ell (\omega)}.
\end{equation}
It is important to also recall that the solutions of the homogeneous Zerilli-Moncrief and Regge-Wheeler equations (\ref{H_ZMetRW_equation}) are related by the Chandrasekhar-Detweiler transformation \cite{Chandrasekhar:1975zza,Chandrasekhar:1985kt}
\begin{widetext}
\begin{equation}\label{ChandraDet_transf_PHI}
\left[\Lambda (\Lambda+2) \mp i (12 M\omega) \right] \phi^{(e/o)}_{\omega\ell} =\left[\Lambda (\Lambda+2) + \frac{72 M^2}{r (\Lambda r+6M)} f(r) \pm 12M f(r) \frac{d}{dr} \right] \phi^{(o/e)}_{\omega\ell}.
\end{equation}
\end{widetext}
As a consequence, the coefficients $A^{(\pm,{p})}_\ell (\omega)$ satisfy the relations
\begin{subequations}\label{ChandraDet_transf}
\begin{equation}\label{ChandraDet_transf_a}
A^{(-,e)}_\ell (\omega)=A^{(-,o)}_\ell (\omega)
\end{equation}
and
\begin{eqnarray}\label{ChandraDet_transf_b}
& & [\Lambda (\Lambda +2) - i  (12M\omega) ] A^{(+,e)}_\ell (\omega) \nonumber \\
& & \qquad \qquad = [\Lambda (\Lambda +2) + i  (12M\omega) ] A^{(+,o)}_\ell (\omega)
\end{eqnarray}
\end{subequations}
and the definition (\ref{Matrix_S}) provides
\begin{eqnarray}\label{Matrix_S_rel}
& & [\Lambda (\Lambda +2) - i  (12M\omega) ]  S^{(e)}_{\ell}(\omega) \nonumber \\
& &  \qquad \qquad = [\Lambda (\Lambda +2) + i  (12M\omega) ] S^{(o)}_{\ell}(\omega).
\end{eqnarray}

\subsection{CAM representation of the scattering amplitudes $f^{+}(\omega,x)$ and $f^{-}(\omega,x)$}
\label{SecIIb}

\subsubsection{Sommerfeld-Watson representation of the scattering amplitudes}
\label{SecIIb1}

\begin{figure}%[h!]
 \includegraphics[scale=0.50]{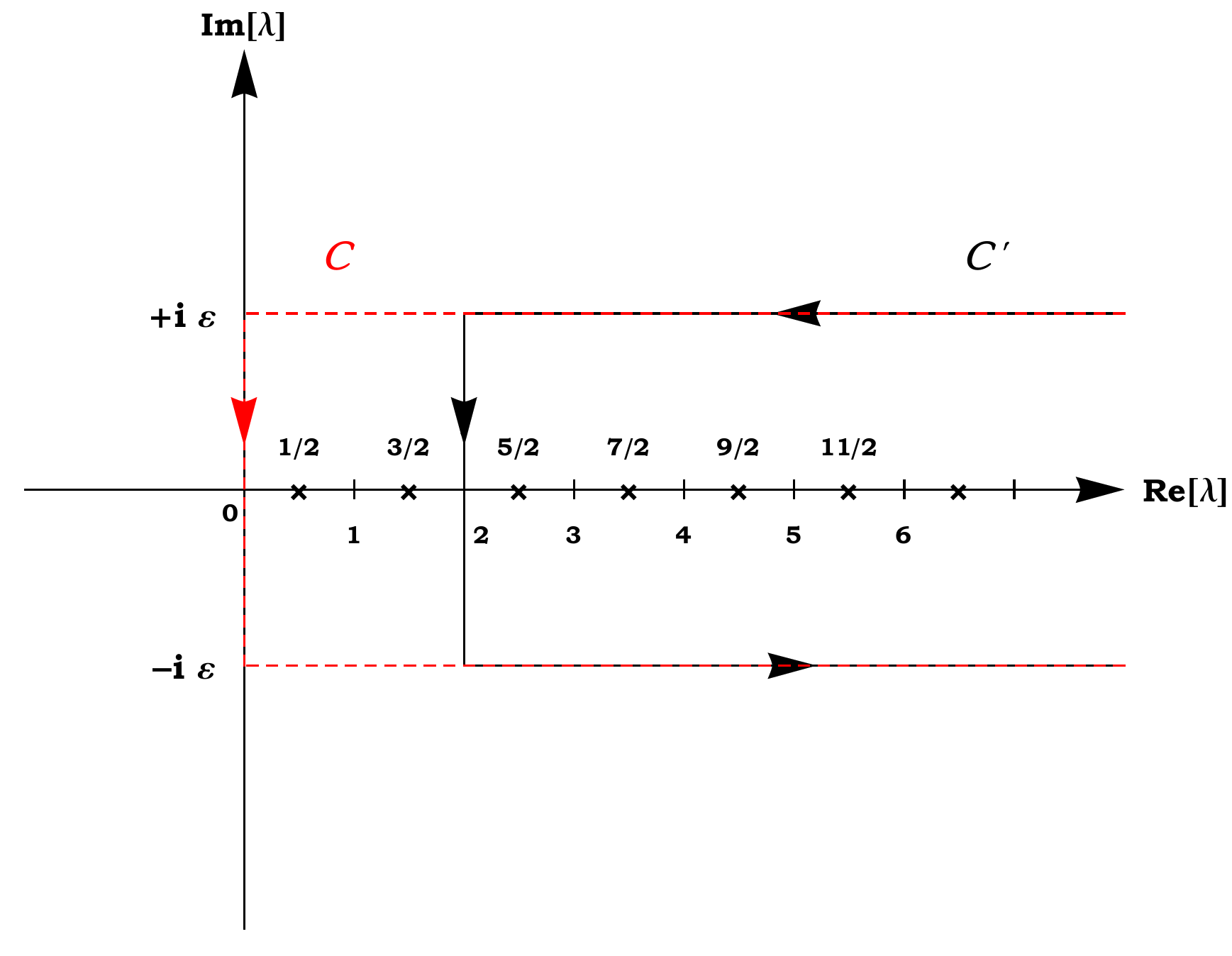}
\caption{\label{Contour_Shift_s2} Integration contours in the CAM plane: ${\cal C'}$ is associated with the scattering amplitudes (\ref{SW_GW_Scattering_amp}) and ${\cal C}$ with the scattering amplitudes (\ref{SW_GW_Scattering_amp_prov1}).}
\end{figure}

By means of the Sommerfeld-Watson transformation \cite{Watson18,Sommerfeld49,Newton:1982qc} which permits us to write
\begin{equation}\label{SWT_gen}
\sum_{\ell=2}^{+\infty} (-1)^\ell F(\ell)= \frac{i}{2} \int_{\cal C'} d\lambda \, \frac{F(\lambda -1/2)}{\cos (\pi \lambda)}
\end{equation}
for a function $F$ without any singularities on the real $\lambda$ axis, we can replace in Eq.~(\ref{GW_Scattering_amp_tilde}) the discrete sum over the ordinary angular momentum $\ell$ by a contour integral in the
complex $\lambda$ plane (i.e., in the complex $\ell$ plane with $\lambda = \ell +1/2$). By noting that $P_\ell (x)=(-1)^\ell P_\ell (-x)$, we obtain
\begin{eqnarray}\label{SW_GW_Scattering_amp}
& & {\widetilde f}^{\pm}(\omega,x) = \frac{1}{2 \omega}  \int_{\cal C'} d\lambda \, \frac{\lambda}{(\lambda^2-1/4)(\lambda^2-9/4)\cos (\pi \lambda)} \nonumber \\
&&     \times \left[ \frac{1}{2} \left(S^{(e)}_{\lambda -1/2}(\omega) \pm S^{(o)}_{\lambda -1/2}(\omega) \right) - \left( \frac{1 \pm 1}{2} \right) \right] P_{\lambda -1/2} (-x).\nonumber \\
&&
\end{eqnarray}
In Eqs.~(\ref{SWT_gen}) and (\ref{SW_GW_Scattering_amp}), we have taken ${\cal C'}=]+\infty +i\epsilon,2+i\epsilon] \cup [2+i\epsilon,2-i\epsilon] \cup [2-i\epsilon, +\infty -i\epsilon[$
with $\epsilon \to 0_+$ (see Fig.~\ref{Contour_Shift_s2}).  We can recover (\ref{GW_Scattering_amp_tilde}) from (\ref{SW_GW_Scattering_amp}) by using
Cauchy's residue theorem and by noting that the poles of the integrand in
(\ref{SW_GW_Scattering_amp}) are the zeros of $\cos (\pi \lambda)$ that are enclosed into ${\cal C'}$, i.e., the semi-integers $\lambda = \ell + 1/2$ with $\ell \in \mathbb{N}- \lbrace{0, 1\rbrace}$. It should be recalled that, in Eq.~(\ref{SW_GW_Scattering_amp}), the Legendre function of first kind $P_{\lambda -1/2} (z)$ denotes the analytic extension of the Legendre polynomials $P_\ell (z)$. It is defined in terms of hypergeometric functions by \cite{AS65}
\begin{equation}\label{Def_ext_LegendreP}
P_{\lambda -1/2} (z) = F[1/2-\lambda,1/2+\lambda;1;(1-z)/2].
\end{equation}
Similarly, in Eq.~(\ref{SW_GW_Scattering_amp}), $S^{(e)}_{\lambda -1/2} (\omega)$ and $S^{(o)}_{\lambda -1/2} (\omega)$ denote the analytic extensions of the matrices $S^{(e)}_\ell (\omega)$ and $S^{(o)}_\ell (\omega)$. We shall briefly mention some of their properties.

\begin{figure}%[h!]
 \includegraphics[scale=0.50]{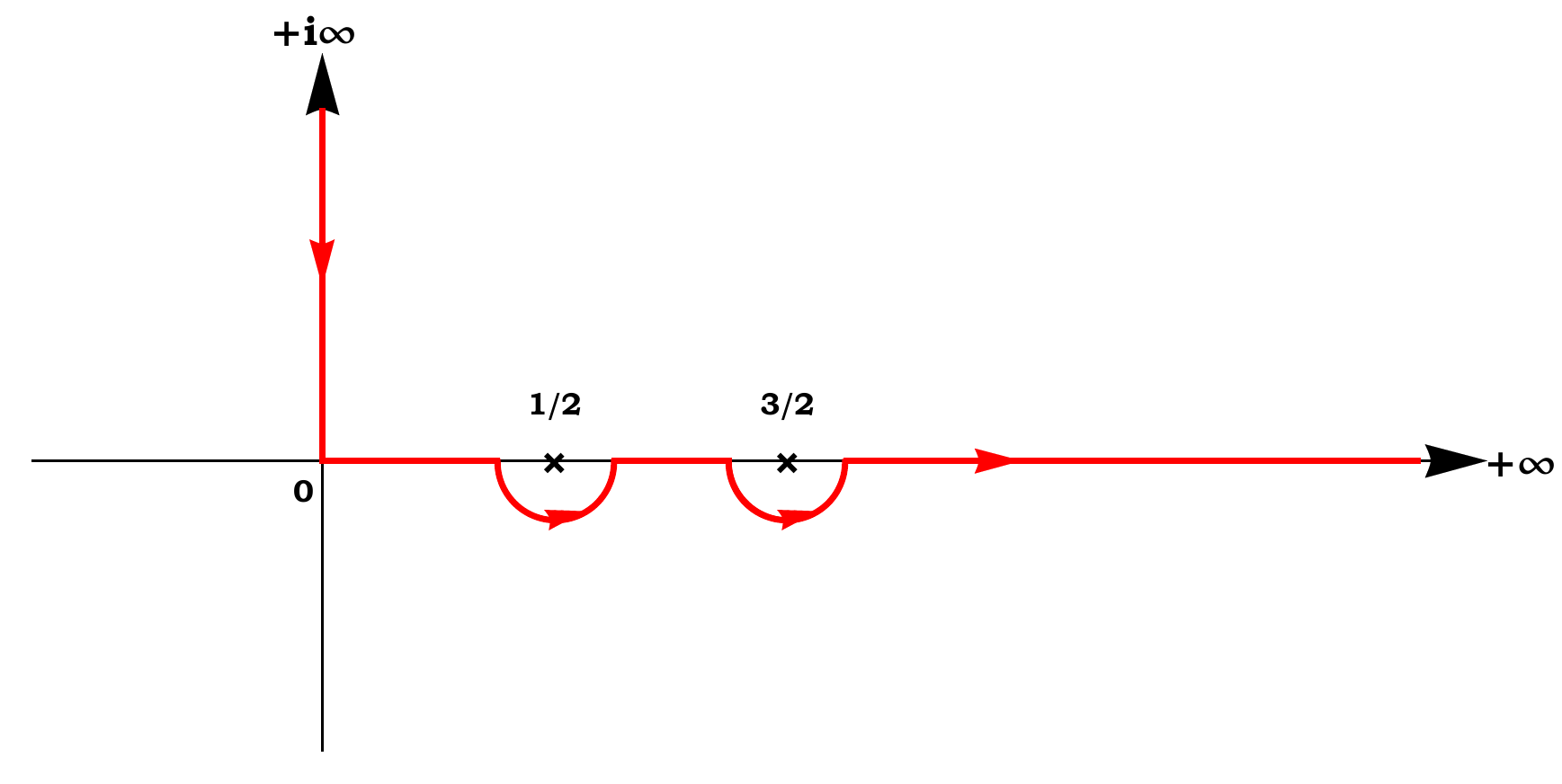}
\caption{\label{Contour_integ_de_fond_s_2} The path of integration in the CAM plane which defines the background integrals $f^{\pm}_\text{\tiny{B}} (\omega,x)$ given by (\ref{CAM_GW_Scattering_amp_decomp}).}
\end{figure}

\subsubsection{Analytic extensions of the matrices $S^{(e)}_{\ell}(\omega)$ and $S^{(o)}_{\ell}(\omega)$ in the CAM plane}
\label{SecIIb2}

We can consider that the analytic extension of the matrix $S^{(p)}_{\ell}(\omega)$ is given by [see Eq.~(\ref{Matrix_S})]
\begin{equation}\label{Matrix_S_CAM}
  S^{(p)}_{\lambda -1/2}(\omega) =  e^{i(\lambda + 1/2)\pi} \, \frac{A_{\lambda -1/2}^{(+,p)}(\omega)}{A_{\lambda -1/2}^{(-,p)}(\omega)}
\end{equation}
where the complex amplitudes $A^{(-,p)}_{\lambda -1/2} (\omega)$ and  $A^{(+,p)}_{\lambda -1/2} (\omega)$ are defined from the analytic extensions of the modes $\phi_{\omega \ell}^{\mathrm {in} \, {(p)}}$, i.e., from the functions $\phi_{\omega ,\lambda -1/2}^{\mathrm {in} \, {(p)}}$ solutions of the problem (\ref{H_ZMetRW_equation})-(\ref{bc_in}) where we now replace $\ell$ by $\lambda -1/2$. It is worth noting that the Chandrasekhar-Detweiler relations (\ref{ChandraDet_transf}) lead, in the CAM plane, to
\begin{subequations}\label{ChandraDet_transf_CAM}
\begin{equation}\label{ChandraDet_transf_CAM_a}
A^{(-,e)}_{\lambda-1/2} (\omega)=A^{(-,o)}_{\lambda-1/2} (\omega)
\end{equation}
and
\begin{eqnarray}\label{ChandraDet_transf_CAM_b}
& & [(\lambda^2-1/4)(\lambda^2-9/4) - i  (12M\omega) ] A^{(+,e)}_{\lambda-1/2} (\omega) \nonumber \\
& & \qquad  =[(\lambda^2-1/4)(\lambda^2-9/4) + i  (12M\omega) ] A^{(+,o)}_{\lambda-1/2} (\omega) \nonumber \\
& &
\end{eqnarray}
\end{subequations}
and, as a consequence, that the definition (\ref{Matrix_S_CAM}) provides
\begin{eqnarray}\label{Matrix_S_rel_CAM}
& & [(\lambda^2-1/4)(\lambda^2-9/4) - i  (12M\omega) ]  S^{(e)}_{\lambda-1/2}(\omega) \nonumber \\
& & \qquad  =[(\lambda^2-1/4)(\lambda^2-9/4) + i  (12M\omega) ] S^{(o)}_{\lambda-1/2}(\omega). \nonumber \\
& &
\end{eqnarray}

Here, with the deformation of the contour ${\cal C'}$ in mind, it is important to note the symmetry property
\begin{equation}\label{Matrix_S_CAM_symm}
  e^{ i \pi \lambda} \, S^{(p)}_{-\lambda -1/2}(\omega) =  e^{- i \pi \lambda} \, S^{(p)}_{\lambda -1/2}(\omega)
\end{equation}
which can be easily obtained from the definition (\ref{Matrix_S_CAM}) (see also Ref.~\cite{Andersson:1994rk}). It is also important to remark that, due to the relation (\ref{ChandraDet_transf_CAM_a}), the poles of the matrices $S^{(e)}_{\lambda -1/2}(\omega)$ and $S^{(o)}_{\lambda -1/2}(\omega)$ in the complex $\lambda$ plane (i.e., the so-called Regge poles) are identical. They lie in the first and third quadrants of the CAM plane, symmetrically distributed with respect to the origin $O$ of this plane, and the poles $\lambda_n(\omega)$ with $n=1, 2, 3, \dots  $ lying in the first quadrant satisfy
\begin{equation}\label{PR_def_Am}
A^{(-,e)}_{\lambda_n(\omega)-1/2} (\omega)=A^{(-,o)}_{\lambda_n(\omega)-1/2} (\omega)=0.
\end{equation}

In the following, the residues of the matrices $S^{(e)}_{\lambda-1/2}(\omega)$ and $S^{(o)}_{\lambda-1/2}(\omega)$ at these poles will play a central role. They are defined by
\begin{equation}\label{residues_RP}
r^{(p)}_n(\omega)=e^{i\pi [\lambda_n(\omega)+1/2]} \left[ \frac{A_{\lambda -1/2}^{(+,p)}(\omega)}{\frac{d}{d \lambda}A_{\lambda -1/2}^{(-,p)}(\omega)}\right]_{\lambda=\lambda_n(\omega)}.
\end{equation}
It is moreover worth noting that, due to the Chandrasekhar-Detweiler relations (\ref{ChandraDet_transf_CAM}), we have
\begin{eqnarray}\label{resS_rel_CAM}
& & \left\lbrace[\lambda_n(\omega)^2-1/4][\lambda_n(\omega)^2-9/4] - i  (12M\omega) \right\rbrace  r^{(e)}_n(\omega) \nonumber \\
& & \quad  =\left\lbrace[\lambda_n(\omega)^2-1/4][\lambda_n(\omega)^2-9/4] + i  (12M\omega)  \right\rbrace r^{(o)}_n(\omega). \nonumber \\
& &
\end{eqnarray}

\subsubsection{Modification of the contour defining the Sommerfeld-Watson representation of the scattering amplitudes}
\label{SecIIb3}

With the aim of collecting the Regge poles in mind, it is necessary to move the contour ${\cal C'}$ to the left so that it coincides with the contour ${\cal C}=]+\infty +i\epsilon, +i\epsilon] \cup
[+i\epsilon,-i\epsilon] \cup [-i\epsilon, +\infty -i\epsilon[$ (see Fig.~\ref{Contour_Shift_s2}). Here, we extend our treatment of scattering of electromagnetic waves \cite{Folacci:2019cmc}. However, we then introduce two spurious double poles: the first one at $\lambda=1/2$ (i.e., at $\ell=0$) which comes from the term $1/[(\lambda-1/2)\cos (\pi \lambda)]$ and the second one at $\lambda=3/2$ (i.e., at $\ell=1$) which comes from the term $1/[(\lambda-3/2)\cos (\pi \lambda)]$. It is necessary to remove the associated residue contributions and we obtain
\begin{widetext}
\begin{eqnarray}\label{SW_GW_Scattering_amp_prov1}
& & 2 \omega {\widetilde f}^{\pm}(\omega,x) =  \int_{\cal C} d\lambda \, \frac{\lambda}{(\lambda^2-1/4)(\lambda^2-9/4)\cos (\pi \lambda)} \left[ \frac{1}{2} \left(S^{(e)}_{\lambda -1/2}(\omega) \pm S^{(o)}_{\lambda -1/2}(\omega) \right) - \left( \frac{1 \pm 1}{2} \right) \right] P_{\lambda -1/2} (-x) \nonumber \\
&&    - 2i\pi \lim_{\lambda \to 1/2} \frac{d}{d\lambda} \left\lbrace (\lambda-1/2)^2 \times \frac{\lambda}{(\lambda^2-1/4)(\lambda^2-9/4)\cos (\pi \lambda)} \left[ \frac{1}{2} \left(S^{(e)}_{\lambda -1/2} (\omega) \pm S^{(o)}_{\lambda -1/2}(\omega) \right) - \left( \frac{1 \pm 1}{2} \right) \right] P_{\lambda -1/2} (-x) \right\rbrace  \nonumber \\
&&    - 2i\pi \lim_{\lambda \to 3/2} \frac{d}{d\lambda} \left\lbrace (\lambda-3/2)^2 \times \frac{\lambda}{(\lambda^2-1/4)(\lambda^2-9/4)\cos (\pi \lambda)} \left[ \frac{1}{2} \left(S^{(e)}_{\lambda -1/2}(\omega) \pm S^{(o)}_{\lambda -1/2}(\omega) \right) - \left( \frac{1 \pm 1}{2} \right) \right] P_{\lambda -1/2} (-x)\right\rbrace. \nonumber \\
& &
\end{eqnarray}
The terms neutralizing the contributions of the spurious poles can be evaluated explicitly by using, in particular,
\begin{subequations} \label{der_FuncLeg}
\begin{equation}\label{der_FuncLeg_1}
\left[\frac{d}{d\lambda} P_{\lambda-1/2}(-x)\right]_{\lambda=1/2} = \ln \left( \frac{1-x}{2} \right)
\end{equation}
and
\begin{equation}\label{der_FuncLeg_3}
\left[\frac{d}{d\lambda} P_{\lambda-1/2}(-x)\right]_{\lambda=3/2} = -(1+x) -x \ln \left( \frac{1-x}{2} \right)
\end{equation}
\end{subequations}
and we have
\begin{eqnarray}\label{SW_GW_Scattering_amp_prov2}
& & 2 \omega {\widetilde f}^{\pm}(\omega,x) =  \int_{\cal C} d\lambda \, \frac{\lambda}{(\lambda^2-1/4)(\lambda^2-9/4)\cos (\pi \lambda)} \left[ \frac{1}{2} \left(S^{(e)}_{\lambda -1/2}(\omega) \pm S^{(o)}_{\lambda -1/2}(\omega) \right) - \left( \frac{1 \pm 1}{2} \right) \right] P_{\lambda -1/2} (-x) \nonumber \\
&&    \qquad\qquad + \frac{i}{2} \left[ \frac{1}{2} \left(S^{(e)}_{1}(\omega) \pm S^{(o)}_{1}(\omega) \right) - \left( \frac{1 \pm 1}{2} \right) \right] \left[(1+x) + x \ln \left( \frac{1-x}{2} \right)\right]   \nonumber \\
&&   \qquad\qquad - \frac{i}{2} \left[ \frac{1}{2} \left(S^{(e)}_{0}(\omega) \pm S^{(o)}_{0}(\omega) \right) - \left( \frac{1 \pm 1}{2} \right) \right] \ln \left( \frac{1-x}{2} \right)   \nonumber \\
&&    \qquad\qquad + \frac{i}{2} \left\lbrace  \left[ \frac{1}{2} \frac{d}{d\lambda}  \left(S^{(e)}_{\lambda -1/2}(\omega) \pm S^{(o)}_{\lambda -1/2}(\omega) \right)  \right]_{\lambda=3/2}
-\frac{7}{6}\left[ \frac{1}{2} \left(S^{(e)}_{1}(\omega) \pm S^{(o)}_{1}(\omega) \right) - \left( \frac{1 \pm 1}{2} \right) \right]  \right\rbrace x  \nonumber \\
&&    \qquad\qquad
+ \mathrm{terms \,\, independent \,\, of \,\,} x.
\end{eqnarray}
The terms $S^{(p)}_{0}(\omega)$, $S^{(p)}_{1}(\omega)$ and $[d/d\lambda \, S^{(p)}_{\lambda -1/2}(\omega)]_{\lambda=3/2}$ appearing in Eq.~(\ref{SW_GW_Scattering_amp_prov2}) can be numerically determined by solving the problem (\ref{H_ZMetRW_equation})-(\ref{bc_in}). However, they do not contribute necessarily to the scattering amplitudes (\ref{GW_Scattering_amp}). Indeed, by applying the differential operators ${\widehat {\cal L}}^{\pm}_{x}$ on (\ref{SW_GW_Scattering_amp_prov2}), we can show that
\begin{eqnarray}\label{SW_GW_Scattering_amp_fin}
& & f^{\pm}(\omega,x) = {\widehat {\cal L}}^{\pm}_{x} \left\lbrace \frac{1}{2 \omega} \int_{\cal C} d\lambda \, \frac{\lambda}{(\lambda^2-1/4)(\lambda^2-9/4)\cos (\pi \lambda)} \right. \nonumber \\
& & \left. \qquad\qquad \times \left[ \frac{1}{2} \left(S^{(e)}_{\lambda -1/2}(\omega) \pm S^{(o)}_{\lambda -1/2}(\omega) \right) - \left( \frac{1 \pm 1}{2} \right) \right] P_{\lambda -1/2} (-x) \right\rbrace  + f_\text{\tiny{SP}}^{\pm}(\omega,x)
\end{eqnarray}
\end{widetext}
where the amplitudes $f_\text{\tiny{SP}}^{\pm} (\omega,x)$ denote the contributions of the terms introduced in order to neutralize the spurious double poles at $\lambda=1/2$ and $\lambda=3/2$. These ``shift corrections'' are given by
\begin{subequations} \label{SW_GW_Scattering_amp_compl}
\begin{eqnarray}
& & f_\text{\tiny{SP}}^{+}(\omega,x)  =0, \label{SW_GW_Scattering_amp_compl_a} \\
& & f_\text{\tiny{SP}}^{-}(\omega,x)  = -\frac{3 i}{\omega} \left[S^{(e)}_{1}(\omega) - S^{(o)}_{1}(\omega) \right]  \frac{1}{(1-x)^2} \nonumber \\
& & \qquad + \frac{i}{\omega} \left[S^{(e)}_{0}(\omega) - S^{(o)}_{0}(\omega) \right]  \frac{2+x}{(1-x)^2}. \label{SW_GW_Scattering_amp_compl_b}
\end{eqnarray}
\end{subequations}
Here it should be noted that the last expression could be slightly simplified by using the fact that [see Eq.~(\ref{Matrix_S_rel})]
\begin{equation}\label{S0_and_S1_rel}
S^{(e)}_\ell(\omega) + S^{(o)}_\ell (\omega)=0  \quad \mathrm{for} \quad \ell = 0 \quad \mathrm{and} \quad 1.
\end{equation}
It is also interesting to note that only the helicity-reversing scattering amplitude $f^{-}(\omega,x)$ is affected by the shift of the integration contour. The helicity-preserving scattering amplitude $f^{+}(\omega,x)$, just like the scattering amplitude associated with electromagnetic waves \cite{Folacci:2019cmc}, is not altered.

\subsubsection{CAM representation of the scattering amplitudes}
\label{SecIIb4}

We now deform the contour ${\cal C}$ in Eq.~(\ref{SW_GW_Scattering_amp_fin}) in order to collect, by using Cauchy's residue theorem, the Regge pole contributions. This is achieved by following, {\it mutatis mutandis}, the approach developed in Ref.~\cite{Folacci:2019cmc} (see more particularly Sec.~IIB3 and Fig.~1 of this previous article). As we have already noted in this work, this must be done very carefully and, in particular, we must deal with the contributions coming from the quarter circles at infinity with great caution. By using (\ref{Matrix_S_CAM_symm}) and the relation \cite{AS65}
\begin{equation}\label{prop_ext_LegendreP}
P_{-\lambda -1/2} (z) = P_{\lambda -1/2} (z)
\end{equation}
we obtain
\begin{equation}\label{CAM_GW_Scattering_amp_tot}
f^{\pm}(\omega,x) = f^{\pm}_\text{\tiny{B}} (\omega,x) +  f^{\pm}_\text{\tiny{RP}} (\omega,x) +f_\text{\tiny{SP}}^{\pm}(\omega,x)
\end{equation}
where
\begin{subequations}\label{CAM_GW_Scattering_amp_decomp}
\begin{equation}\label{CAM_GW_Scattering_amp_decomp_Background}
f^{\pm}_\text{\tiny{B}} (\omega,x) = f^{\pm}_\text{\tiny{B},\tiny{Re}} (\omega,x)+f^{\pm}_\text{\tiny{B},\tiny{Im}} (\omega,x)
\end{equation}
with
\begin{eqnarray}\label{CAM_GW_Scattering_amp_decomp_Background_a}
& & f^{\pm}_\text{\tiny{B},\tiny{Re}} (\omega,x) = {\widehat {\cal L}}^{\pm}_{x} \left\lbrace \frac{1}{2\pi \omega} \int_{{\cal C}_{-}} d\lambda \, \frac{\lambda}{(\lambda^2-1/4)(\lambda^2-9/4)}  \right. \nonumber \\
& &   \left. \phantom{\int_{{\cal C}_{-}}}  \times \left[S^{(e)}_{\lambda -1/2}(\omega) \pm S^{(o)}_{\lambda -1/2}(\omega) \right] \,Q_{\lambda -1/2}(x +i0)  \right\rbrace
\end{eqnarray}
and
\begin{eqnarray}\label{CAM_GW_Scattering_amp_decomp_Background_b}
& & f^{\pm}_\text{\tiny{B},\tiny{Im}} (\omega,x) = {\widehat {\cal L}}^{\pm}_{x} \left\lbrace \frac{1}{2\pi \omega} \int_{+i\infty}^0 d\lambda \, \frac{\lambda}{(\lambda^2-1/4)(\lambda^2-9/4)}  \right. \nonumber \\
& &   \left. \phantom{\int_{\infty}^0}  \times \left[S^{(e)}_{\lambda -1/2}(\omega) \pm S^{(o)}_{\lambda -1/2}(\omega) \right] \,Q_{\lambda -1/2}(x +i0)  \right\rbrace
\end{eqnarray}
\end{subequations}
is a background integral contribution (here we have  ${\cal C}_{-}=
[0,-i\epsilon] \cup [-i\epsilon, +\infty -i\epsilon[$
with $\epsilon \to 0_+$) and where
\begin{eqnarray}\label{CAM_GW_Scattering_amp_decomp_RP}
& & f^{\pm}_\text{\tiny{RP}} (\omega,x) = {\widehat {\cal L}}^{\pm}_{x} \left\lbrace
-\frac{i \pi}{2\omega}    \sum_{n=1}^{+\infty}    \right. \nonumber \\
&&  \left. \phantom{\sum_{n=1}^{+\infty}} \frac{ \lambda_n(\omega) \left[ r^{(e)}_n(\omega) \pm r^{(o)}_n(\omega) \right]}{[\lambda_n(\omega)^2-1/4][\lambda_n(\omega)^2-9/4] \cos[\pi \lambda_n(\omega)]} \right. \nonumber \\
&&  \left. \phantom{\sum_{n=1}^{+\infty}} \times    P_{\lambda_n(\omega) -1/2} (-x)  \right\rbrace
\end{eqnarray}
is a sum over the Regge poles lying in the first quadrant of the CAM plane involving the residues of the matrices $S^{(e)}_{\lambda-1/2}(\omega)$ and $S^{(o)}_{\lambda-1/2}(\omega)$ at these poles [see Eq.~(\ref{residues_RP})].
In Eqs.~(\ref{CAM_GW_Scattering_amp_decomp_Background_a}) and (\ref{CAM_GW_Scattering_amp_decomp_Background_b}), we have introduced the Legendre function of the second kind $Q_{\lambda -1/2}(z)$ and used the relation \cite{AS65}
\begin{eqnarray}\label{Legendre_function_second_kind}
& & Q_{\lambda -1/2}(x +i0) = \frac{\pi}{2 \cos(\pi \lambda)} \left[ P_{\lambda -1/2}(-x) \right. \nonumber \\
& & \qquad \qquad \left. - e^{-i\pi (\lambda-1/2)} P_{\lambda -1/2}(+x )\right].
\end{eqnarray}
Of course, Eqs.~(\ref{CAM_GW_Scattering_amp_tot})-(\ref{CAM_GW_Scattering_amp_decomp_RP}) and (\ref{SW_GW_Scattering_amp_compl}) provide exact representations of the scattering amplitudes $f^{\pm} (\omega, x)$ equivalent to the initial partial wave expansions defined by Eqs.~(\ref{GW_Scattering_diff})-(\ref{GW_Scattering_OpDiff}). From these CAM representations, we can extract the contributions denoted by $f^{\pm}_\text{\tiny{RP}} (\omega, x)$ given by (\ref{CAM_GW_Scattering_amp_decomp_RP}) which, as sums over Regge poles, are only approximations of the scattering amplitudes $f^{\pm} (\omega, x)$ and which can provide us with an approximation of the differential scattering cross section (\ref{GW_Scattering_diff}).

\subsubsection{Important remarks concerning background integrals}
\label{SecIIb5}

It is important to note that the path of integration associated with the background integrals $f^{\pm}_\text{\tiny{B}} (\omega,x) = f^{\pm}_\text{\tiny{B},\tiny{Re}} (\omega,x)+f^{\pm}_\text{\tiny{B},\tiny{Im}} (\omega,x)$ defined by (\ref{CAM_GW_Scattering_amp_decomp}) is a continuous one running down first the positive imaginary axis and then running along ${\cal C}_{-}$, i.e., slightly below the positive real axis. The branch ${\cal C}_{-}$ cannot be deformed in order to coincide exactly with the positive real axis. In fact, it can be deformed taking account the singularities of the integrand in the right-hand side of (\ref{CAM_GW_Scattering_amp_decomp_Background_a}). Indeed, at first sight, this integrand has a simple pole at $\lambda=1/2$ and another one at $\lambda=3/2$ and we must avoid these two poles by moving along semicircles of radius $\epsilon$ with $\epsilon \to 0_+$ lying in the lower complex $\lambda$ plane. In Fig.~\ref{Contour_integ_de_fond_s_2}, we have displayed the path of integration we shall now consider to define the background integrals $f^{\pm}_\text{\tiny{B}} (\omega,x)$.

We can then evaluate the background integral contributions $f^{\pm}_\text{\tiny{B},\tiny{Re}} (\omega,x)$ given by (\ref{CAM_GW_Scattering_amp_decomp_Background_a}) using again Cauchy's theorem. We have
\begin{widetext}
\begin{eqnarray}\label{CAM_EM_Scattering_amp_decomp_Background_a_simplification}
& & \int_{{\cal C}_{-}} d\lambda \, \frac{\lambda}{(\lambda^2-1/4)(\lambda^2-9/4)} \left[S^{(e)}_{\lambda -1/2}(\omega) \pm S^{(o)}_{\lambda -1/2}(\omega) \right] Q_{\lambda -1/2}(\cos \theta +i0) \nonumber \\
& & \qquad = \mathrm{P.V.} \int_0^{+\infty} d\lambda \, \frac{\lambda}{(\lambda^2-1/4)(\lambda^2-9/4)} \left[S^{(e)}_{\lambda -1/2}(\omega) \pm S^{(o)}_{\lambda -1/2}(\omega) \right] Q_{\lambda -1/2}(\cos \theta +i0) \nonumber \\
& & \qquad \phantom{=}  +i \pi \lim_{\lambda \to 1/2}  \left[ (\lambda-1/2) \times  \frac{\lambda}{(\lambda^2-1/4)(\lambda^2-9/4)} \left[S^{(e)}_{\lambda -1/2}(\omega) \pm S^{(o)}_{\lambda -1/2}(\omega) \right] Q_{\lambda -1/2}(\cos \theta +i0) \right]
\nonumber \\
& & \qquad \phantom{=}  +i \pi \lim_{\lambda \to 3/2}  \left[ (\lambda-3/2) \times  \frac{\lambda}{(\lambda^2-1/4)(\lambda^2-9/4)} \left[S^{(e)}_{\lambda -1/2}(\omega) \pm S^{(o)}_{\lambda -1/2}(\omega) \right] Q_{\lambda -1/2}(\cos \theta +i0) \right]
\end{eqnarray}
where $\mathrm{P.V.}$ denotes the Cauchy principal value symbol associated with the treatment of the singularities of the integrand at $\lambda=1/2$ and $\lambda=3/2$. The two residue contributions in (\ref{CAM_EM_Scattering_amp_decomp_Background_a_simplification}) can be evaluated explicitly by using the expression (\ref{Legendre_function_second_kind}) as well as Eqs.~(\ref{der_FuncLeg_1}) and (\ref{der_FuncLeg_3}). We then obtain
\begin{eqnarray}\label{CAM_GW_Scattering_amp_decomp_Background_a_simplification_bis}
& & \int_{{\cal C}_{-}} d\lambda \, \frac{\lambda}{(\lambda^2-1/4)(\lambda^2-9/4)} \left[S^{(e)}_{\lambda -1/2}(\omega) \pm S^{(o)}_{\lambda -1/2}(\omega) \right] Q_{\lambda -1/2}(\cos \theta +i0) \nonumber \\
& & \qquad = \mathrm{P.V.} \int_0^{+\infty} d\lambda \, \frac{\lambda}{(\lambda^2-1/4)(\lambda^2-9/4)} \left[S^{(e)}_{\lambda -1/2}(\omega) \pm S^{(o)}_{\lambda -1/2}(\omega) \right] Q_{\lambda -1/2}(\cos \theta +i0) \nonumber \\
& & \qquad \phantom{=}  + \frac{i \pi}{8} \left[S^{(e)}_{0}(\omega) \pm S^{(o)}_{0}(\omega) \right] \left[ \ln \left( \frac{1-x}{2} \right) - \ln \left( \frac{1+x}{2} \right) \right]
\nonumber \\
& & \qquad \phantom{=} - \frac{i \pi}{8} \left[S^{(e)}_{1}(\omega) \pm S^{(o)}_{1}(\omega) \right] \left[ x \ln \left( \frac{1-x}{2} \right) - x \ln \left( \frac{1+x}{2}\right) +i\pi x  \right] \nonumber \\
& & \qquad \phantom{=} + \mathrm{terms \,\, independent \,\, of \,\,} x.
\end{eqnarray}
By finally applying the differential operators ${\widehat {\cal L}}^{\pm}_{x}$ on this relation and by taking into account (\ref{S0_and_S1_rel}), we can show that
\begin{eqnarray}\label{CAM_GW_Scattering_amp_decomp_Background_fpm_axis}
& & f^{\pm}_\text{\tiny{B},\tiny{Re}} (\omega,x) = {\widehat {\cal L}}^{\pm}_{x} \left\lbrace \mathrm{P.V.} \frac{1}{2\pi \omega} \int_0^{+\infty} d\lambda \, \frac{\lambda}{(\lambda^2-1/4)(\lambda^2-9/4)} \right. \nonumber \\
& & \qquad\qquad  \left. \phantom{\int_0^{+\infty}} \times \left[S^{(e)}_{\lambda -1/2}(\omega) \pm S^{(o)}_{\lambda -1/2}(\omega) \right] \,Q_{\lambda -1/2}(x +i0)  \right\rbrace - \frac{1}{2} f_\text{\tiny{SP}}^{\pm}(\omega,x)
\end{eqnarray}
with $f_\text{\tiny{SP}}^{\pm}(\omega,x)$ given by (\ref{SW_GW_Scattering_amp_compl}).
\end{widetext}

It is interesting to note that, in fact, the integrand of the function $f^{+}_\text{\tiny{B},\tiny{Re}} (\omega,x)$ given by (\ref{CAM_GW_Scattering_amp_decomp_Background_a}) is regular. Indeed, the divergence of $1/(\lambda-1/2)$ for $\lambda=1/2$ and of $1/(\lambda-3/2)$ for $\lambda=3/2$ is compensated by the vanishing of $S^{(e)}_{\lambda -1/2}(\omega) + S^{(o)}_{\lambda -1/2}(\omega)$ [see Eq.~(\ref{S0_and_S1_rel})] and the path ${\cal C}_{-}$ can be deformed in order to coincide exactly with the positive real axis. In other words, when if we define $f^{+}_\text{\tiny{B},\tiny{Re}} (\omega,x)$ by (\ref{CAM_GW_Scattering_amp_decomp_Background_fpm_axis}), it is not necessary to consider the Cauchy principal value of the integral over the real positive axis and we can take for the integration contour defining the background integral $f^{+}_\text{\tiny{B}} (\omega,x) = f^{+}_\text{\tiny{B},\tiny{Re}} (\omega,x)+f^{+}_\text{\tiny{B},\tiny{Im}} (\omega,x)$ given by (\ref{CAM_GW_Scattering_amp_decomp}) the path $]+i \infty,0] \, \cup \, [0,+\infty[$.

\section{Reconstruction of scattering amplitudes and of the differential scattering cross section from Regge pole sums}
\label{SecIII}

In this section, we compare numerically the exact differential scattering cross section defined by (\ref{GW_Scattering_diff})-(\ref{GW_Scattering_amp_tilde}) as well as the helicity-preserving and helicity-reversing scattering amplitudes (\ref{GW_Scattering_amp})-(\ref{GW_Scattering_amp_tilde}) with their CAM representations constructed in Sec.~\ref{SecIIb} and, more particularly, with their Regge pole approximations (\ref{CAM_GW_Scattering_amp_decomp_RP}). This permits us to highlight the benefits of working with Regge pole sums in the short-wavelength regime and the necessity to include, in the long-wavelength regime, the contribution of the background integrals (\ref{CAM_GW_Scattering_amp_decomp}) and of the shift corrections (\ref{SW_GW_Scattering_amp_compl}).

\subsection{Computational methods}
\label{SecIIIa}

In order to numerically construct the scattering amplitudes (\ref{GW_Scattering_amp})-(\ref{GW_Scattering_amp_tilde}), the differential scattering cross section (\ref{GW_Scattering_diff})-(\ref{GW_Scattering_amp_tilde}), the background integrals (\ref{CAM_GW_Scattering_amp_decomp}) as well as the Regge pole sums (\ref{CAM_GW_Scattering_amp_decomp_RP}), we use, {\it mutatis mutandis}, the computational methods that have permitted us, in Ref.~\cite{Folacci:2019cmc}, to revisit from CAM techniques the scattering of scalar and electromagnetic waves by a Schwarzschild BH. We refer the reader to Sec.~IIIA of this previous article but also to Secs.~IIIB and IVA of Ref.~\cite{Folacci:2018sef} for the aspects linked with Regge poles. We note that, due to the long-range nature of the fields propagating on the Schwarzschild BH, the scattering amplitudes (\ref{GW_Scattering_amp})-(\ref{GW_Scattering_amp_tilde}) and the background integrals (\ref{CAM_GW_Scattering_amp_decomp_Background_a}) suffer of a lack of convergence [this is not the case for the background integrals (\ref{CAM_GW_Scattering_amp_decomp_Background_b}) because their integrands vanish exponentially as $\lambda \to +i\infty$]. The methods permitting us to overcome this problem, i.e., to accelerate the convergence of these sums and integrals, are described in the Appendix of Ref.~\cite{Folacci:2019cmc}.  Finally, it should be noted that we have performed all the numerical calculations by using {\it Mathematica} \cite{Mathematica}.

\subsection{Results and comments}
\label{SecIIIb}

\begin{figure*}%[htb]
 \includegraphics[scale=0.60]{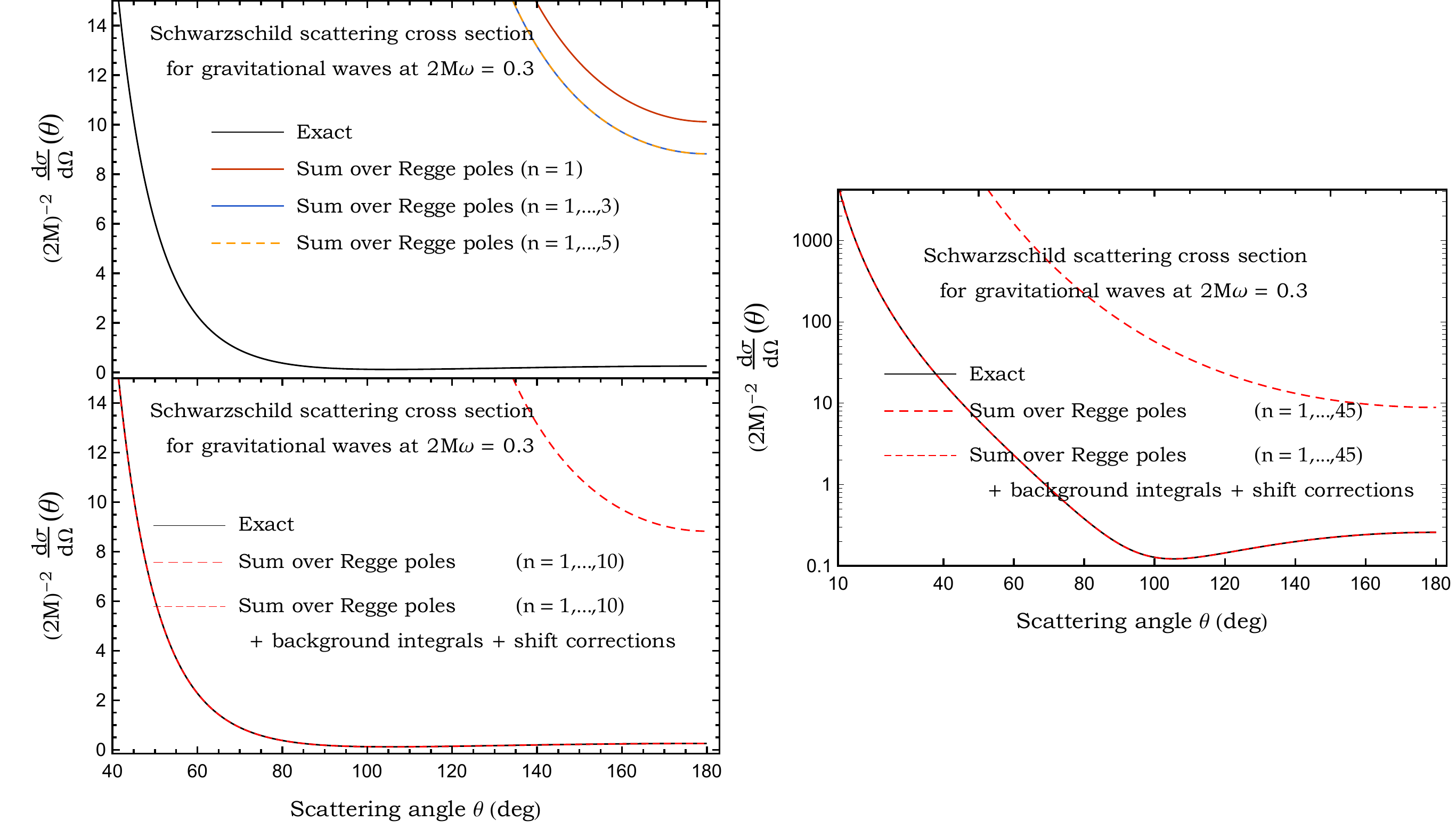}
\caption{\label{S_2_2Mw_03_Exact_vs_CAM} Scattering cross section of a Schwarzschild BH for gravitational waves ($2M\omega=0.3$). We compare the exact cross section defined by (\ref{GW_Scattering_diff})-(\ref{GW_Scattering_amp_tilde}) with its Regge pole approximation constructed from (\ref{CAM_GW_Scattering_amp_decomp_RP}). In addition, we emphasize the role of the background integrals (\ref{CAM_GW_Scattering_amp_decomp}) and of the shift corrections (\ref{SW_GW_Scattering_amp_compl}).}
\end{figure*}

\begin{figure*}%[htb]
 \includegraphics[scale=0.60]{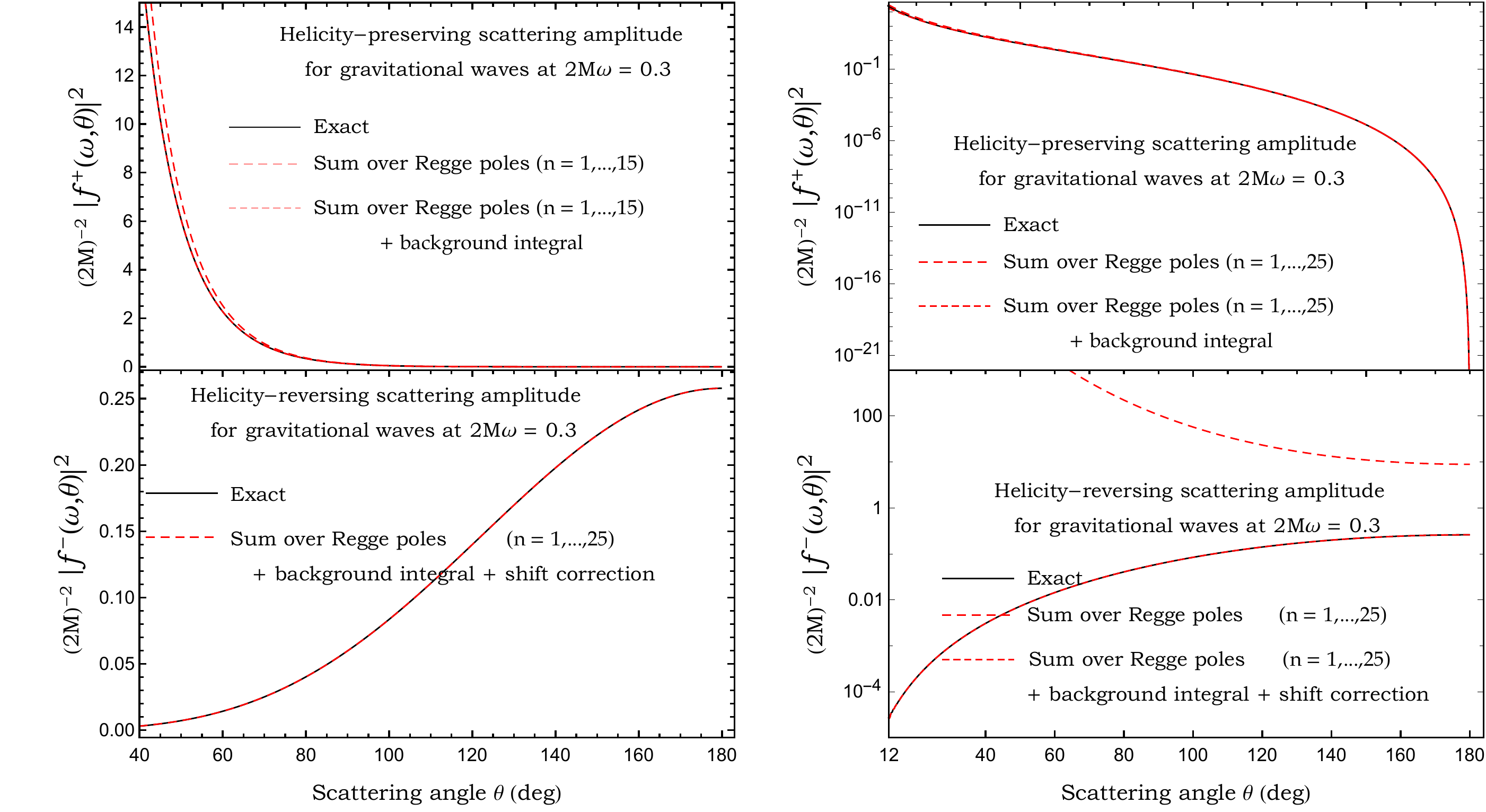}
\caption{\label{S_2_2Mw_03_Exact_vs_CAM_fplus_fmoins} Scattering cross section of a Schwarzschild BH for gravitational waves ($2M\omega=0.3$). Here we consider the helicity-preserving and helicity-reversing scattering amplitudes and we compare the exact results (\ref{GW_Scattering_amp})-(\ref{GW_Scattering_amp_tilde}) with the corresponding Regge pole approximations (\ref{CAM_GW_Scattering_amp_decomp_RP}). In addition, we emphasize the role of the background integrals (\ref{CAM_GW_Scattering_amp_decomp}) and of the shift corrections (\ref{SW_GW_Scattering_amp_compl}).}
\end{figure*}

\begin{figure*}%[htb]
 \includegraphics[scale=0.60]{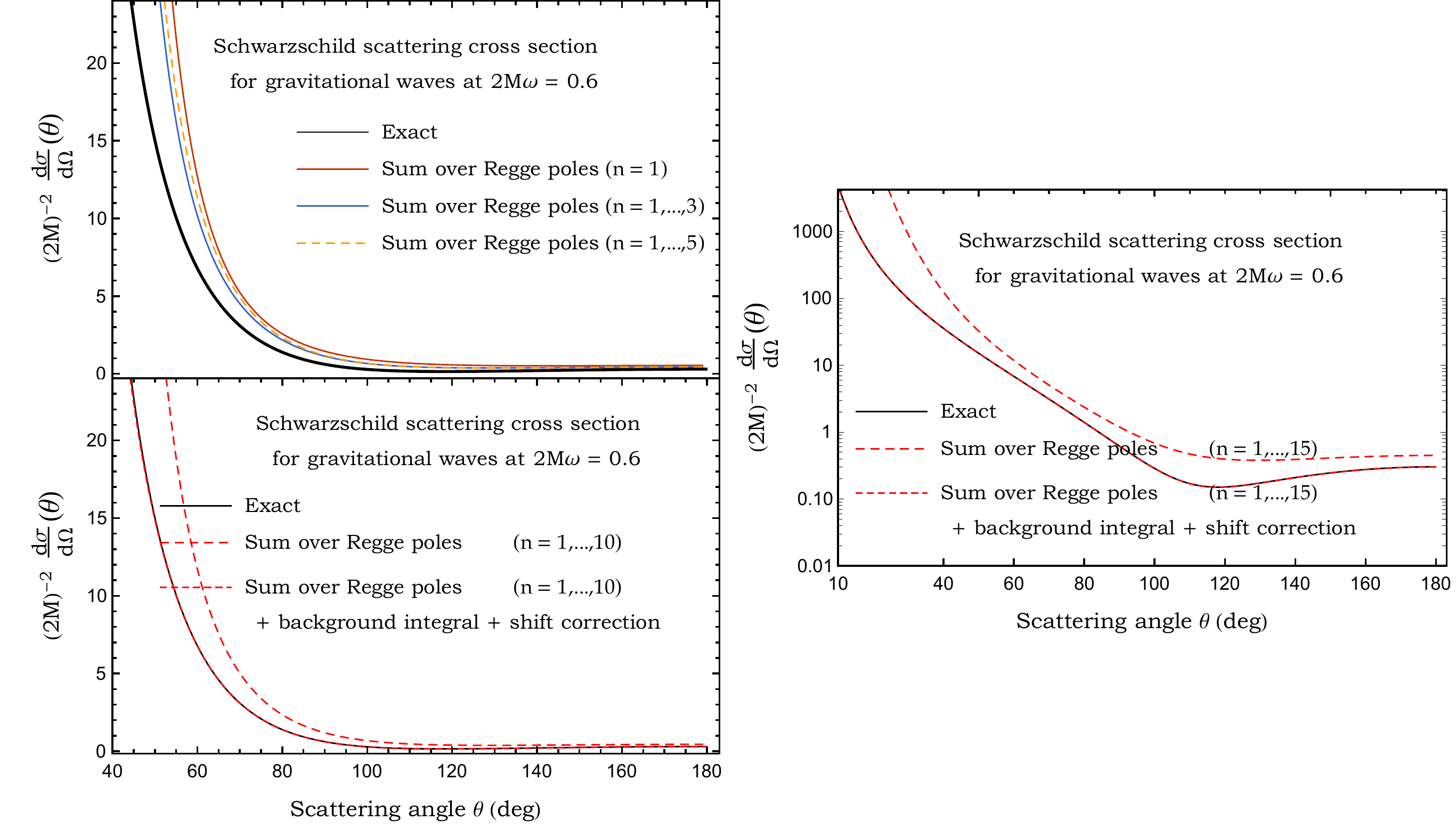}
\caption{\label{S_2_2Mw_06_Exact_vs_CAM} Scattering cross section of a Schwarzschild BH for gravitational waves ($2M\omega=0.6$). We compare the exact cross section defined by (\ref{GW_Scattering_diff})-(\ref{GW_Scattering_amp_tilde}) with its Regge pole approximation constructed from (\ref{CAM_GW_Scattering_amp_decomp_RP}). In addition, we emphasize the role of the background integrals (\ref{CAM_GW_Scattering_amp_decomp}) and of the shift corrections (\ref{SW_GW_Scattering_amp_compl}).}
\end{figure*}

\begin{figure*}%[htb]
 \includegraphics[scale=0.60]{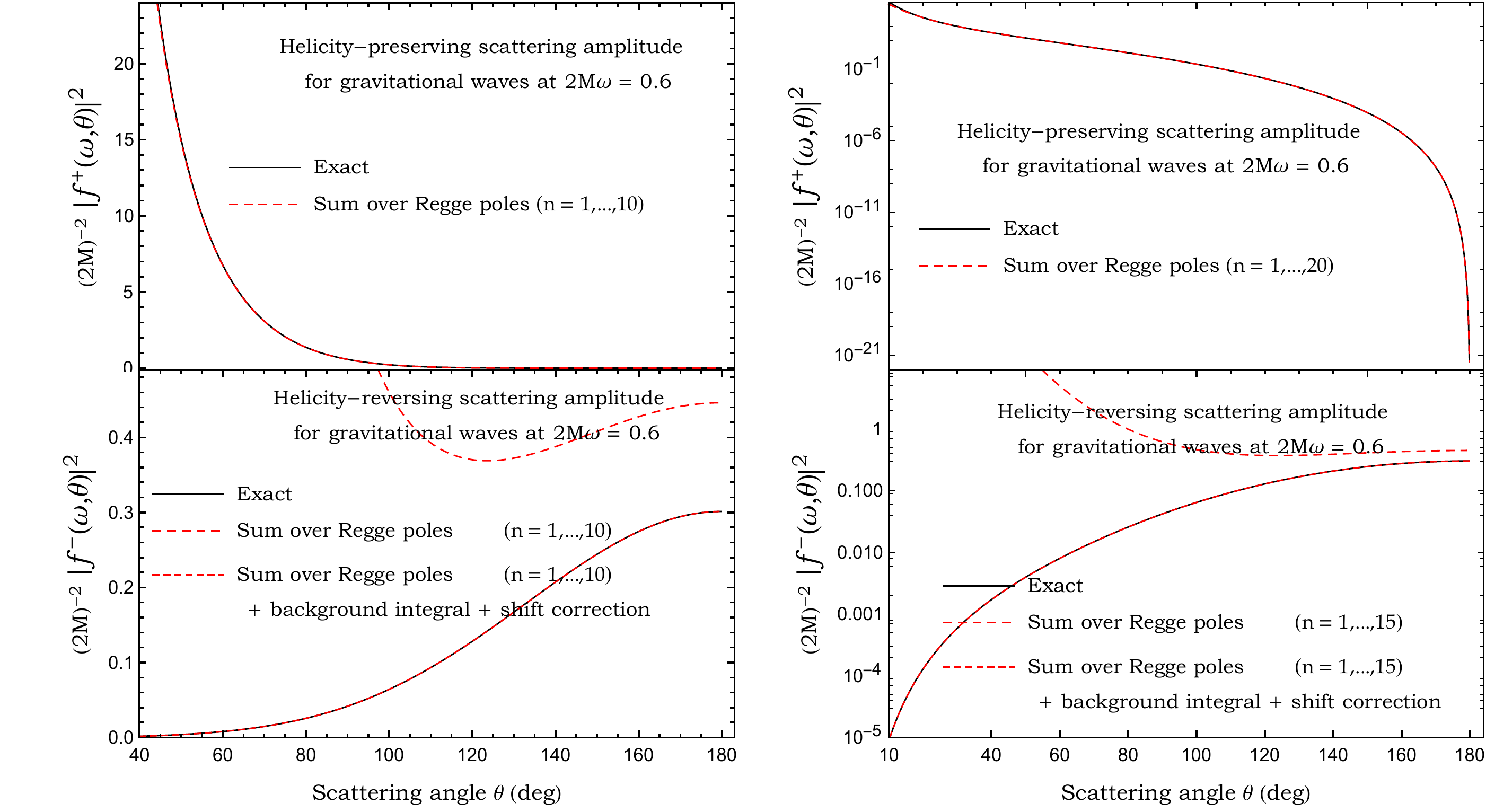}
\caption{\label{S_2_2Mw_06_Exact_vs_CAM_fplus_fmoins} Scattering cross section of a Schwarzschild BH for gravitational waves ($2M\omega=0.6$). Here we consider the helicity-preserving and helicity-reversing scattering amplitudes and we compare the exact results (\ref{GW_Scattering_amp})-(\ref{GW_Scattering_amp_tilde}) with the corresponding Regge pole approximations (\ref{CAM_GW_Scattering_amp_decomp_RP}). In addition, we emphasize the role of the background integrals (\ref{CAM_GW_Scattering_amp_decomp}) and of the shift corrections (\ref{SW_GW_Scattering_amp_compl}).}
\end{figure*}

\begin{figure*}%[htb]
 \includegraphics[scale=0.60]{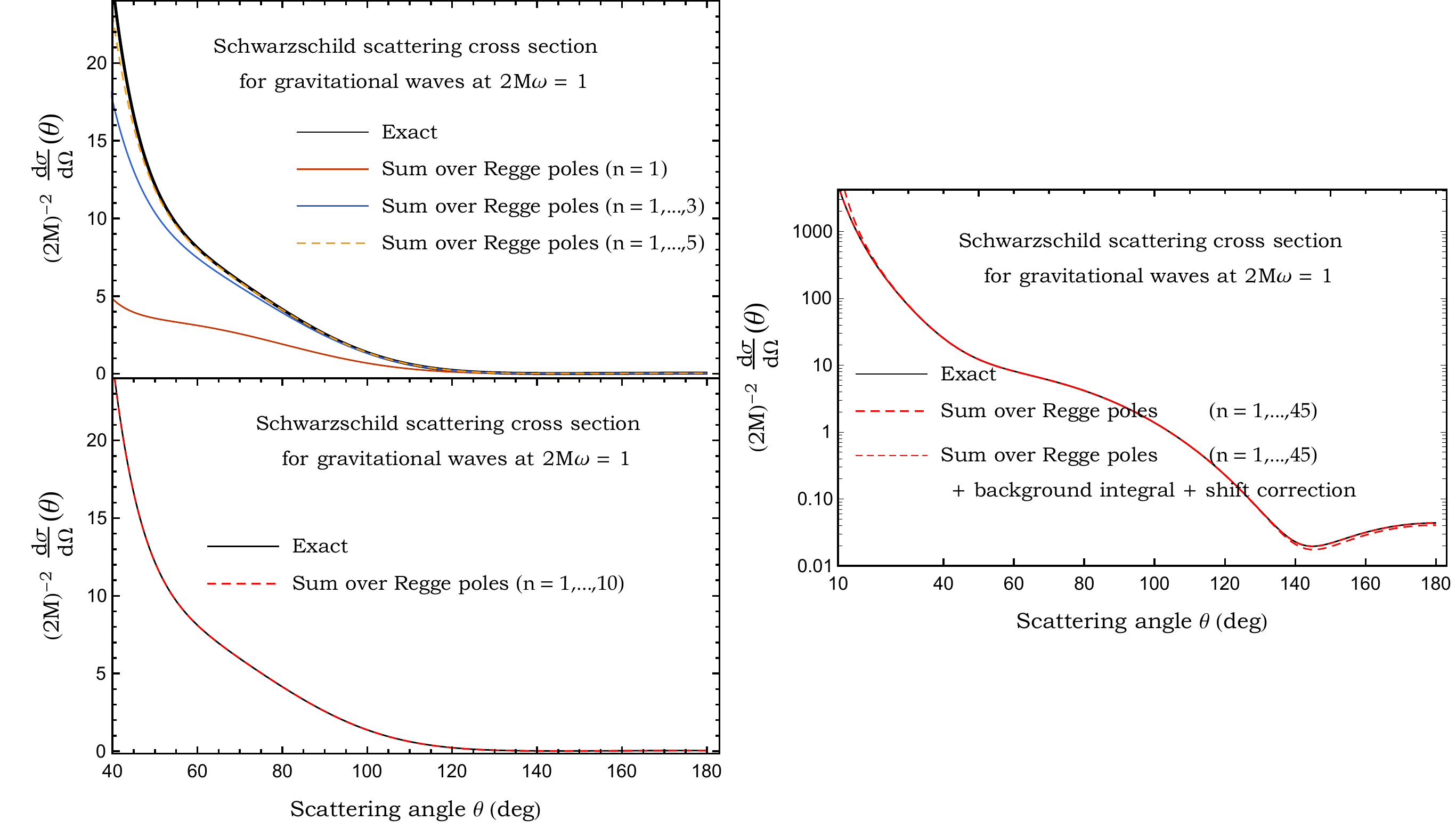}
\caption{\label{S_2_2Mw_1_Exact_vs_CAM} Scattering cross section of a Schwarzschild BH for gravitational waves ($2M\omega=1$). We compare the exact cross section defined by (\ref{GW_Scattering_diff})-(\ref{GW_Scattering_amp_tilde}) with its Regge pole approximation constructed from (\ref{CAM_GW_Scattering_amp_decomp_RP}). In addition, we emphasize the role of the background integrals (\ref{CAM_GW_Scattering_amp_decomp}) and of the shift corrections (\ref{SW_GW_Scattering_amp_compl}).}
\end{figure*}

\begin{figure*}%[htb]
 \includegraphics[scale=0.60]{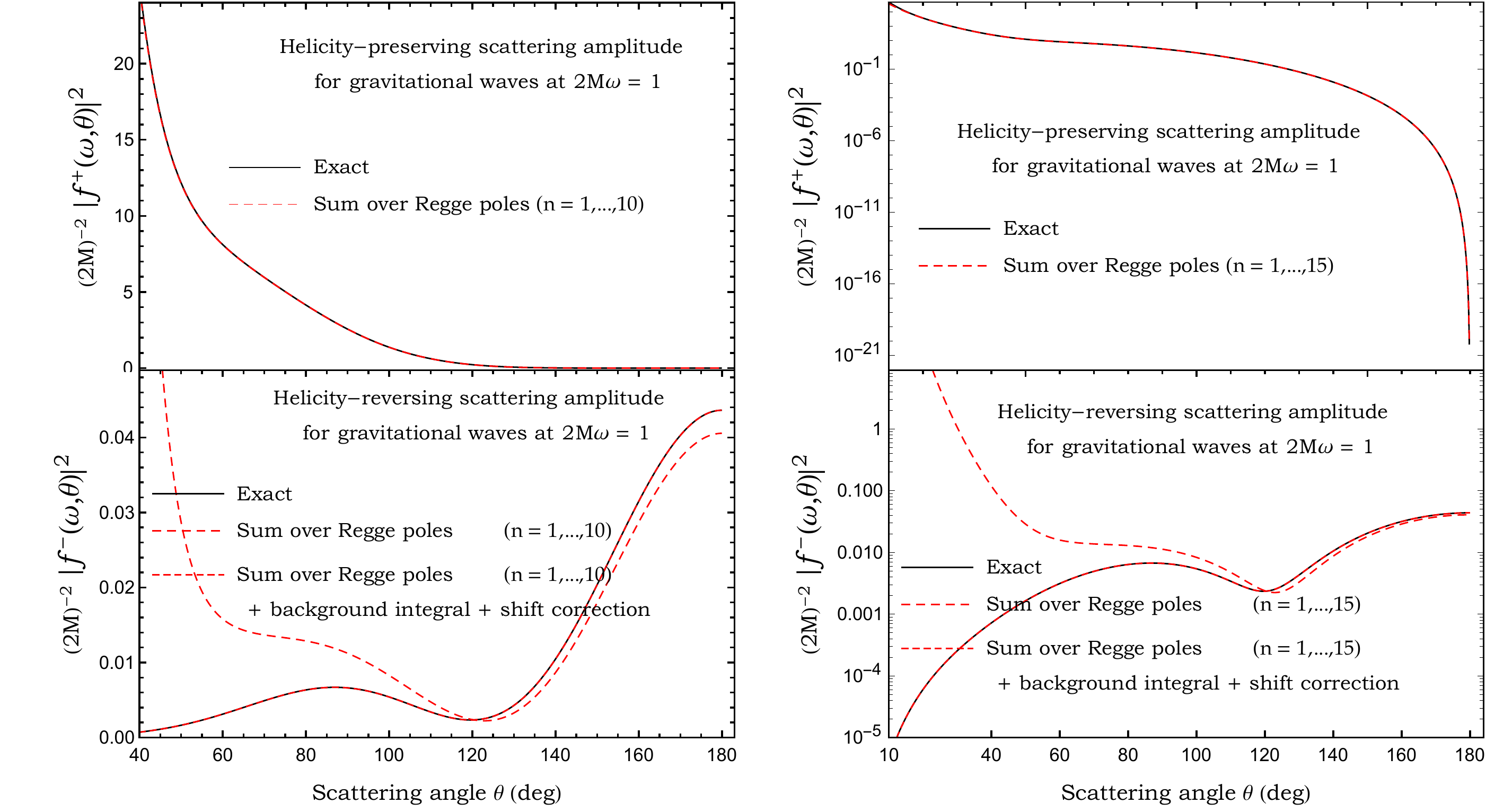}
\caption{\label{S_2_2Mw_1_Exact_vs_CAM_fplus_fmoins} Scattering cross section of a Schwarzschild BH for gravitational waves ($2M\omega=1$). Here we consider the helicity-preserving and helicity-reversing scattering amplitudes and we compare the exact results (\ref{GW_Scattering_amp})-(\ref{GW_Scattering_amp_tilde}) with the corresponding Regge pole approximations (\ref{CAM_GW_Scattering_amp_decomp_RP}). In addition, we emphasize the role of the background integrals (\ref{CAM_GW_Scattering_amp_decomp}) and of the shift corrections (\ref{SW_GW_Scattering_amp_compl}).}
\end{figure*}

\begin{figure*}%[htb]
 \includegraphics[scale=0.60]{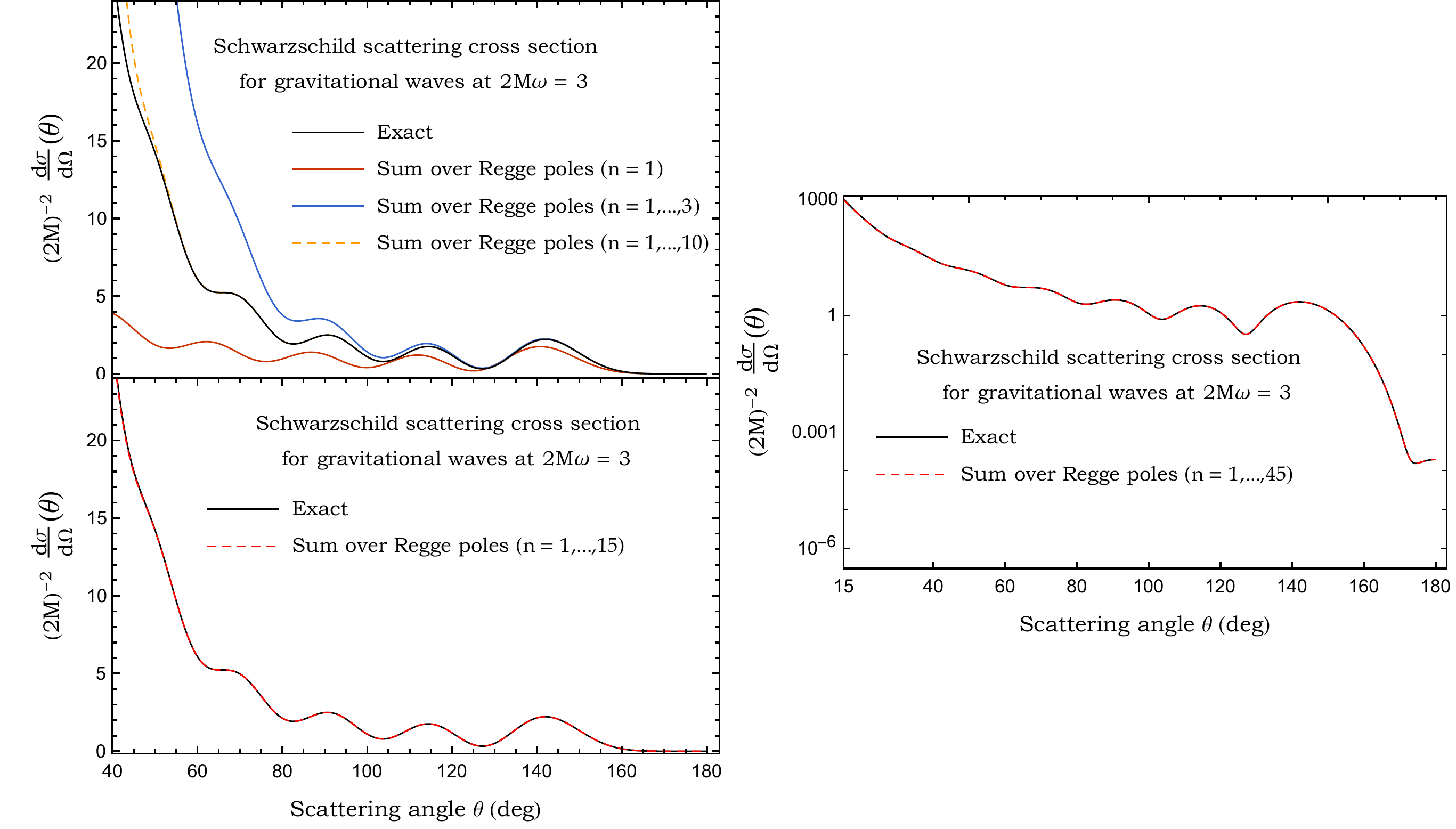}
\caption{\label{S_2_2Mw_3_Exact_vs_CAM} Scattering cross section of a Schwarzschild BH for gravitational waves ($2M\omega=3$). We compare the exact cross section defined by (\ref{GW_Scattering_diff})-(\ref{GW_Scattering_amp_tilde}) with its Regge pole approximation constructed from (\ref{CAM_GW_Scattering_amp_decomp_RP}).}
\end{figure*}

\begin{figure*}%[htb]
 \includegraphics[scale=0.60]{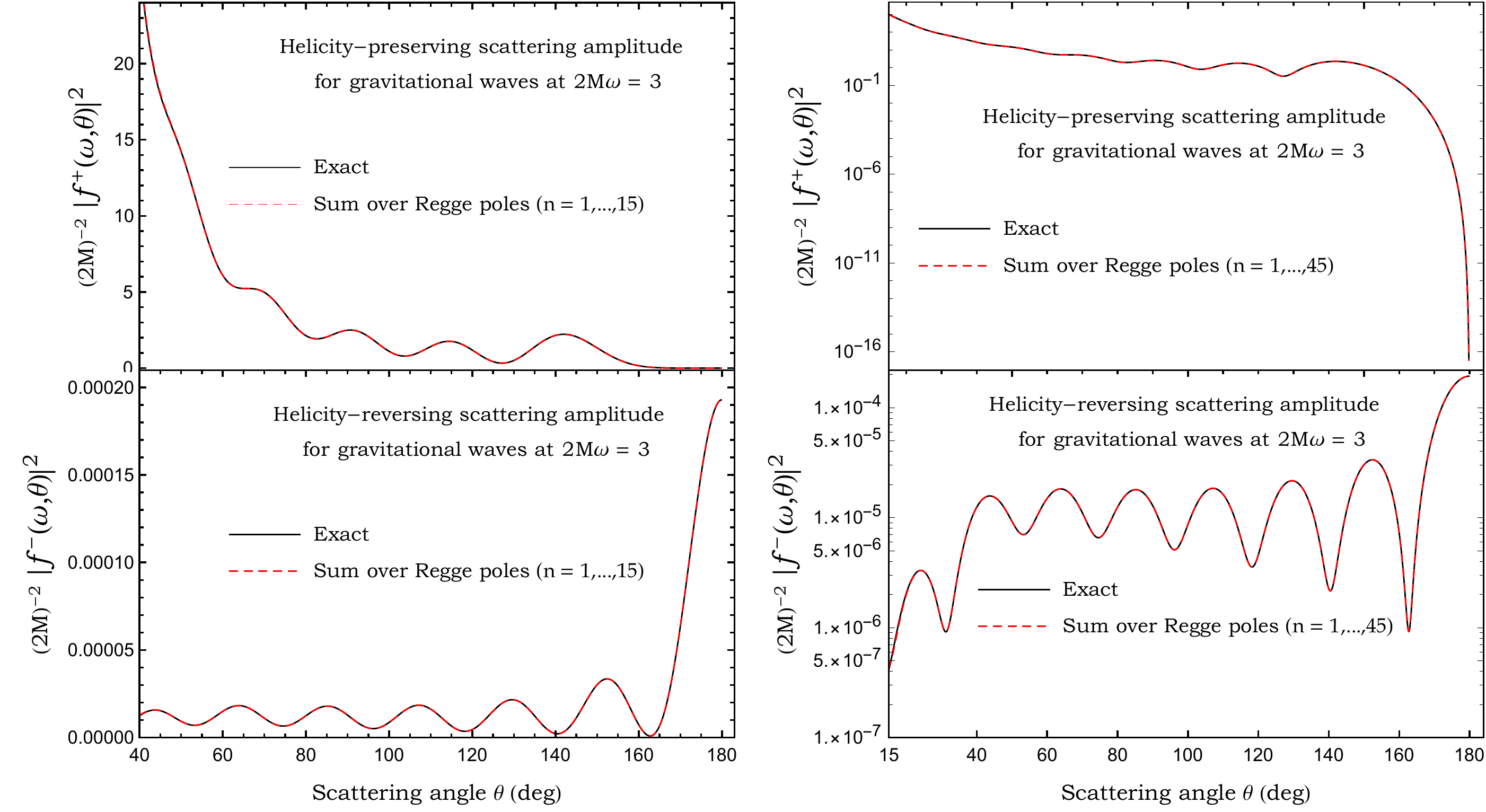}
\caption{\label{S_2_2Mw_3_Exact_vs_CAM_fplus_fmoins} Scattering cross section of a Schwarzschild BH for gravitational waves ($2M\omega=3$). We emphasize the role of the helicity-preserving and helicity-reversing scattering amplitudes and we compare the exact results (\ref{GW_Scattering_amp})-(\ref{GW_Scattering_amp_tilde}) with the corresponding Regge pole approximations (\ref{CAM_GW_Scattering_amp_decomp_RP}).}
\end{figure*}

\begin{figure*}%[htb]
 \includegraphics[scale=0.60]{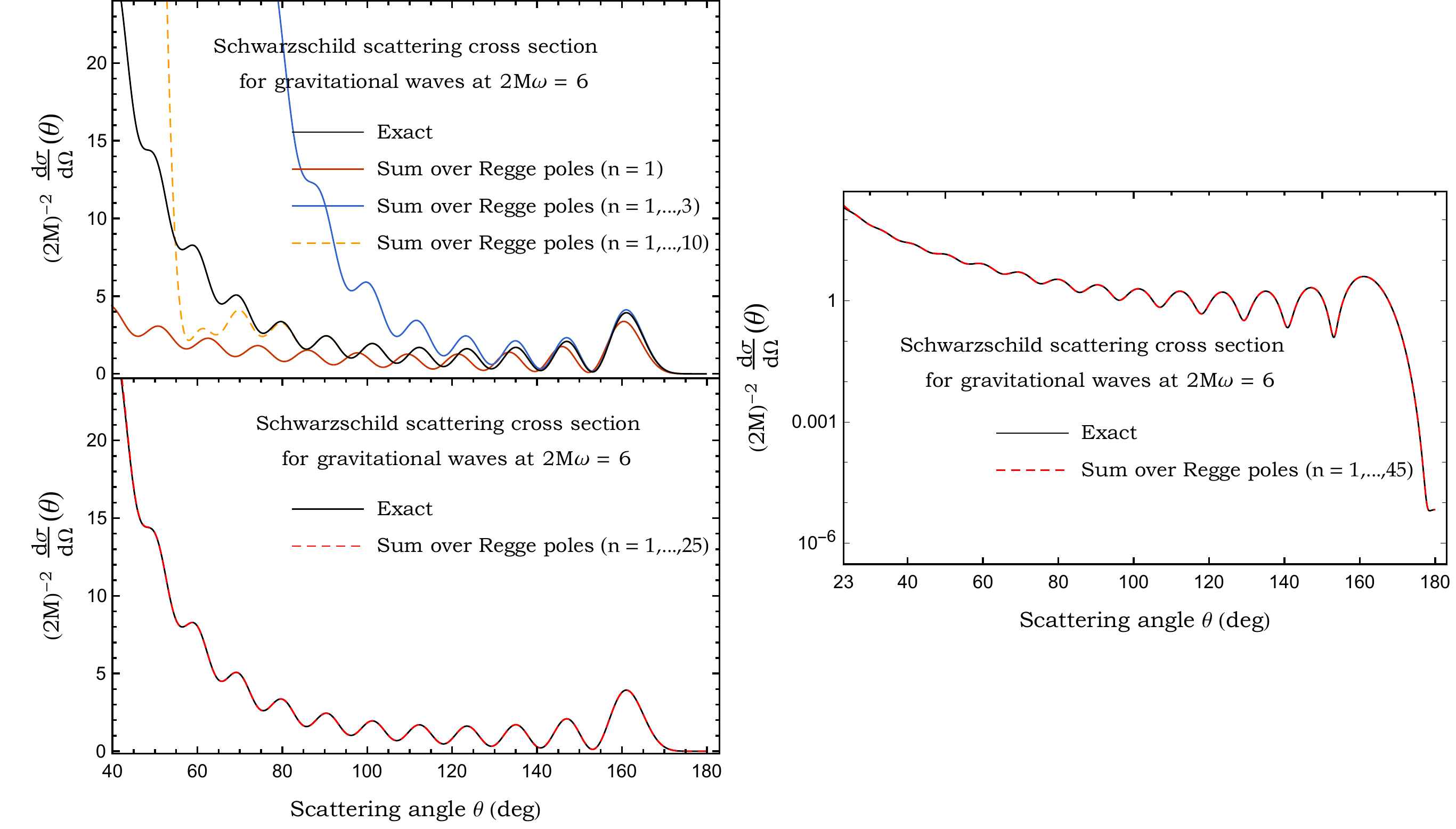}
\caption{\label{S_2_2Mw_6_Exact_vs_CAM} Scattering cross section of a Schwarzschild BH for gravitational waves ($2M\omega=6$). We compare the exact cross section defined by (\ref{GW_Scattering_diff})-(\ref{GW_Scattering_amp_tilde}) with its Regge pole approximation constructed from (\ref{CAM_GW_Scattering_amp_decomp_RP}).}
\end{figure*}

\begin{figure*}%[htb]
 \includegraphics[scale=0.60]{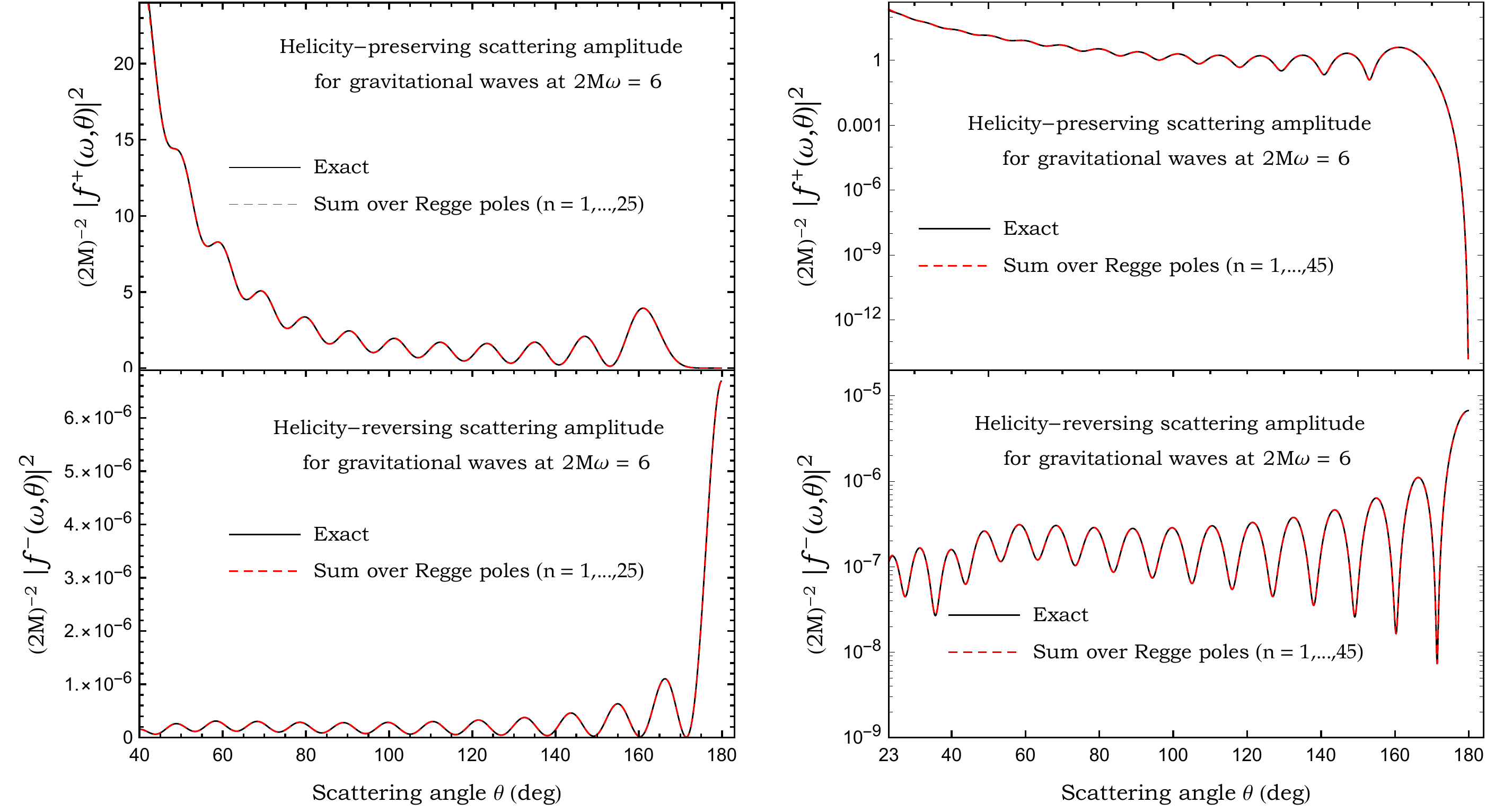}
\caption{\label{S_2_2Mw_6_Exact_vs_CAM_fplus_fmoins} Scattering cross section of a Schwarzschild BH for gravitational waves ($2M\omega=6$). We emphasize the role of the helicity-preserving and helicity-reversing scattering amplitudes and we compare the exact results (\ref{GW_Scattering_amp})-(\ref{GW_Scattering_amp_tilde}) with the corresponding Regge pole approximations (\ref{CAM_GW_Scattering_amp_decomp_RP}).}
\end{figure*}

In Figs.~\ref{S_2_2Mw_03_Exact_vs_CAM}-\ref{S_2_2Mw_6_Exact_vs_CAM_fplus_fmoins}, we have displayed our numerical results. The comparisons of the exact scattering amplitudes and scattering cross sections with their Regge pole approximations and CAM representations have been achieved for the reduced frequencies $2M\omega=$ 0.3, 0.6, 1, 3 and 6 and, for these frequencies, we have provided the lowest Regge poles and the associated residues in Table~\ref{tab:table1}. The higher Regge poles and their residues that have been necessary to obtain some of the results displayed in the figures are available upon request from the authors.

In Figs.~\ref{S_2_2Mw_3_Exact_vs_CAM}-\ref{S_2_2Mw_6_Exact_vs_CAM_fplus_fmoins}, the results correspond to ``high'' reduced frequencies (here we have taken $2M\omega=$ 3 and 6). We can then observe that, in the ``short''-wavelength regime, the Regge pole approximations (\ref{CAM_GW_Scattering_amp_decomp_RP}) involving a small number of Regge poles permit us to describe very well the cross section and the scattering amplitudes for intermediate and large values of the scattering angle and, in particular, the BH glory. Taking into account additional Regge poles improves the Regge pole approximations and we can see that, by summing over a large number of Regge poles, the whole scattering cross section as well as the scattering amplitudes are impressively described, this being valid even for small scattering angles. It is important to note that, in this wavelength regime, it is not necessary to take into account the background integrals and the shift corrections in order to reproduce the scattering amplitudes and the differential scattering cross section. As in the case of the scalar and electromagnetic fields, these contributions are completely negligible for intermediate and large scattering angles. It seems they begin to play a role only for small angles, i.e., for scattering angles $\theta \ll 1/(2M\omega)$.

In Figs.~\ref{S_2_2Mw_03_Exact_vs_CAM}-\ref{S_2_2Mw_1_Exact_vs_CAM_fplus_fmoins}, we focus on the results obtained for low and intermediate reduced frequencies (here we have taken $2M\omega=$ 0.3, 0.6 and 1). We can then observe that, in the long-wavelength regime, the Regge pole approximations (\ref{CAM_GW_Scattering_amp_decomp_RP}) alone do not permit us to reconstruct the scattering amplitudes and the scattering cross sections but that this can be achieved by taking into account the background integral contributions (\ref{CAM_GW_Scattering_amp_decomp}) and the shift corrections (\ref{SW_GW_Scattering_amp_compl}).

\begingroup
\squeezetable
\begin{table}[H]
%\captionsetup{font=small}
\caption{\label{tab:table1} Lowest Regge poles $\lambda_{n}(\omega)$ for gravitational waves and associated residues $r_{n}^{(o)}(\omega)$. We assume $2M=1$. The residues $r_{n}^{(e)}(\omega)$ can be obtained from Eq.~(\ref{resS_rel_CAM}).}
\smallskip
\centering
\begin{ruledtabular}
\begin{tabular}{cccc}

$n$ & $\omega$  & $\lambda_n(\omega)$ & $r^{(o)}_{n}(\omega)$

 \\ \hline
$1$  & $0.3$ &$\phantom{1}1.177734+0.230543 i  $  & $ 0.082479-0.005950 i $   \\
     & $0.6$ &$\phantom{1} 1.677477+0.335822 i  $  & $0.2788875+0.0273350 i $   \\
     & $1$ &$\phantom{1}2.531094+0.411206 i $  & $0.4326579+0.1435142 i $   \\
     & $3$ &$\phantom{1}7.457790+0.486581 i $  & $ -0.2242890+0.8685292 i $   \\
     & $6$ &$ 15.171479+0.496524 i$  & $-0.5321560-1.1687555 i $   \\

$2$   & $0.3$ &$\phantom{1} 0.836492+0.773620 i  $  & $ 0.393716-0.154124 i $   \\
     & $0.6$ &$\phantom{1}1.572789+1.076695 i $  & $0.4447390-0.3967047 i $   \\
     & $1$ &$\phantom{1}2.518269+1.260871 i $  & $ 0.7211612-0.7533923 i $   \\
     & $3$ &$\phantom{1}7.484551+1.457991 i  $  & $4.460584+1.895063 i  $   \\
     & $6$ &$15.188070+1.488756 i$  & $-12.701904+4.668131 i $   \\

$3$  & $0.3$ &$\phantom{1}0.785194+1.696053 i  $  & $ 0.120036-0.314616 i  $   \\
      & $0.6$ &$\phantom{1}1.559438+1.959412 i $  & $ 0.0223704-0.6732064 i $   \\
     & $1$ &$\phantom{1}2.550451+2.151728 i $  & $-0.1658717-1.4015981 i $   \\
     & $3$ &$\phantom{1}7.537660+2.423973 i  $  & $8.960724-10.234313 i $   \\
     & $6$ &$15.220985+2.478570 i $  & $ 13.35540+70.80750 i $   \\

$4$  & $0.3$ &$\phantom{1} 0.900248+2.493219 i  $  & $0.028309-0.277850 i  $   \\
      & $0.6$ &$\phantom{1}1.664569+2.808204 i $  & $-0.2577278-0.5695969 i $   \\
     & $1$ &$\phantom{1}2.649417+3.032799 i  $  & $ -1.043417-1.107841 i $   \\
     & $3$ &$\phantom{1}7.615848+3.380665 i $  & $  -10.02892-25.10010 i$   \\
     & $6$ &$15.269709+3.464449 i $  & $ 259.8284+17.6034 i  $   \\

$5$   & $0.3$ &$\phantom{1}1.042288+3.210140 i  $  & $-0.010991-0.252467 i $   \\
     & $0.6$ &$\phantom{1}1.810963+3.597981 i $  & $ -0.3838196-0.4403928 i  $   \\
     & $1$ &$\phantom{1}2.789282+3.877087 i $  & $-1.437568-0.527774 i $   \\
     & $3$ &$\phantom{1}7.716714+4.324544 i $  & $-42.34029-9.81342 i  $   \\
     & $6$ &$ 15.333503+4.445015 i$  & $286.1595-664.6152 i  $   \\

$6$   & $0.3$ &$\phantom{1}1.187964+3.881018 i  $  & $-0.034909-0.234891 i $   \\
     & $0.6$ &$\phantom{1}1.969737+4.343670 i $  & $-0.4461059-0.3314978 i  $   \\
     & $1$ &$\phantom{1}2.948514+4.684410 i $  & $-1.514306+0.007575 i $   \\
     & $3$ &$\phantom{1}7.836980+5.253021 i $  & $-49.66798+35.28825 i $   \\
     & $6$ &$15.411449+5.419065 i $  & $-1124.069-1219.440 i $   \\

$7$  & $0.3$ &$\phantom{1}1.332119+4.521690 i  $  & $-0.051937-0.221385 i $   \\
     & $0.6$ &$\phantom{1}2.131338+5.057289 i $  & $ -0.4780042-0.2412471 i  $   \\
     & $1$ &$\phantom{1}3.116089+5.460734 i $  & $-1.428013+0.432871 i $   \\
     & $3$ &$\phantom{1}7.973064+6.164682 i $  & $ -16.00828+76.27376 i $   \\
     & $6$ &$15.502498+6.385600 i$  & $-3180.632+810.520 i  $   \\

$8$  & $0.3$ &$\phantom{1} 1.473437+5.140566 i  $  & $ -0.065051-0.210265 i $   \\
      & $0.6$ &$\phantom{1}2.292297+5.746609 i $  & $-0.4930803-0.1653409 i $   \\
     & $1$ &$\phantom{1}3.286697+6.211774 i $  & $-1.262992+0.753379 i  $   \\
     & $3$ &$\phantom{1}8.121586+7.059145 i  $  & $41.25638+84.07005 i  $   \\
     & $6$ &$15.605524+7.343833 i $  & $-1801.097+5504.876 i $   \\

$9$  & $0.3$ &$\phantom{1}1.611665+5.742679 i  $  & $-0.0756224-0.200704 i $   \\
      & $0.6$ &$\phantom{1}2.451286+6.416742 i $  & $-0.4978921-0.1005477 i $   \\
     & $1$ &$\phantom{1}3.457735+6.942014 i $  & $-1.063278+0.987318 i $   \\
     & $3$ &$\phantom{1}8.279630+7.936753 i $  & $ 93.43801+53.60651 i $   \\
     & $6$ &$15.719372+8.293188 i $  & $  5449.250+7574.371 i $   \\

$10$ & $0.3$ &$\phantom{1} 1.746886+6.331313 i  $  & $-0.084397-0.192252 i $   \\
      & $0.6$ &$\phantom{1}2.607816+7.071211 i $  & $-0.4960545-0.0445832 i $   \\
     & $1$ &$\phantom{1}3.627890+7.654858 i $  & $-0.852336+1.152451 i  $   \\
     & $3$ &$\phantom{1}8.444821+8.798278 i $  & $120.10872-1.77950 i  $   \\
     & $6$ &$15.842898+9.233283 i$  & $13938.834+939.263 i  $   \\

%$11$ & $0.3$ &$\phantom{1} 1.879288+6.908754 i $  & $-0.091835-0.184635 i $   \\
%      & $0.6$ &$\phantom{1}2.761755+7.712549 i $  & $-0.4897488+0.0042041 i $   \\
%     & $1$ &$\phantom{1}3.796501+8.352899 i  $  & $ -0.642891+1.263507 i $   \\
%     & $3$ &$\phantom{1}8.615288+9.644699 i $  & $114.81679-62.97254 i $   \\
%     & $6$ &$15.975006+10.163904 i $  & $ 14318.67-13544.52 i  $   \\
%
%$12$ & $0.3$ &$\phantom{1}  2.009084+7.476668 i  $  & $ -0.098235-0.177679 i $   \\
%      & $0.6$ &$\phantom{1}2.913130+8.342648 i $  & $-0.4803759+0.0470473 i  $   \\
%     & $1$ &$\phantom{1}3.963245+9.038149 i $  & $ -0.4418199+1.3321731 i $   \\
%     & $3$ &$\phantom{1}8.789579+10.477066 i $  & $81.98710-113.95343 i $   \\
%     & $6$ &$16.114669+11.084980 i $  & $ 1019.21-26467.35 i  $   \\
%
%$13$  & $0.3$ &$\phantom{1} 2.136480+8.036322 i  $  & $-0.103810-0.171260 i $   \\
%     & $0.6$ &$\phantom{1}3.062037+8.962963 i $  & $-0.4688770+0.0848939 i $   \\
%     & $1$ &$\phantom{1}4.127983+9.712197 i $  & $ -0.2527045+1.3676116 i  $   \\
%     & $3$ &$\phantom{1}8.966580+11.296412 i$  & $31.4975-145.2183 i  $   \\
%     & $6$ &$16.260941+11.996554 i$  & $ -21776.68-26364.70 i$   \\
\end{tabular}
\end{ruledtabular}
\end{table}
\endgroup

\section{BH glory and orbiting oscillations}
\label{SecIV}

\begin{figure*}%[h!]
\centering
 \includegraphics[scale=0.50]{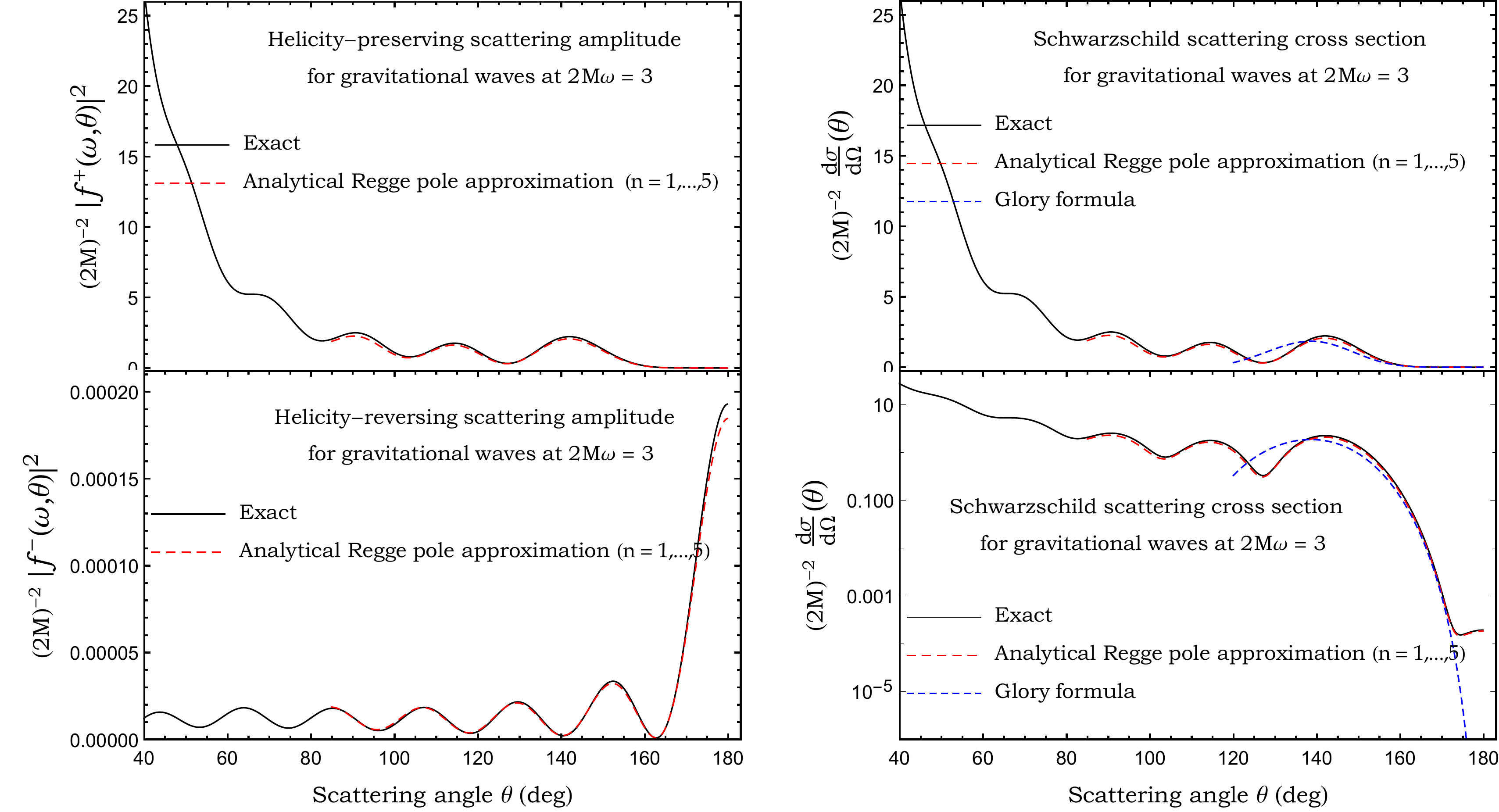}
\caption{\label{RP_approx_2Mw_3_s_2} Scattering amplitudes and scattering cross section of a Schwarzschild BH for gravitational waves ($2M\omega=3$). We compare the exact results given in Sec.~\ref{SecIIa} with those obtained from the analytical Regge pole approximations constructed in Sec.~\ref{SecIV}. We also display the glory cross section described by (\ref{Glory_scattering_formula}).}
\end{figure*}

 \begin{figure*}%[h!]
\centering
 \includegraphics[scale=0.50]{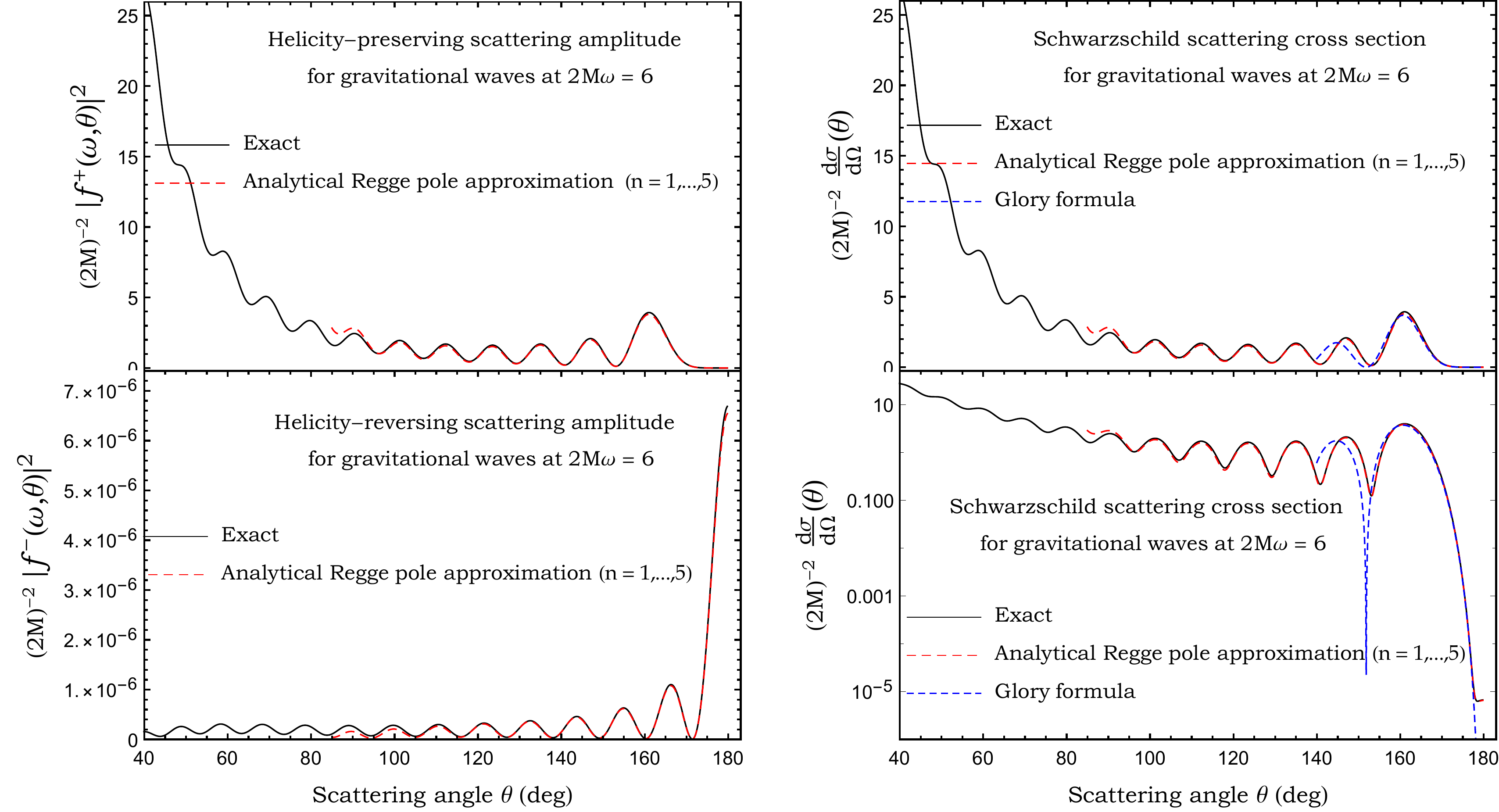}
\caption{\label{RP_approx_2Mw_6_s_2} Scattering amplitudes and scattering cross section of a Schwarzschild BH for gravitational waves ($2M\omega=6$). We compare the exact results given in Sec.~\ref{SecIIa} with those obtained from the analytical Regge pole approximations constructed in Sec.~\ref{SecIV}. We also display the glory cross section described by (\ref{Glory_scattering_formula}).}
\end{figure*}
\vfill

 \begin{figure*}%[h!]
\centering
 \includegraphics[scale=0.50]{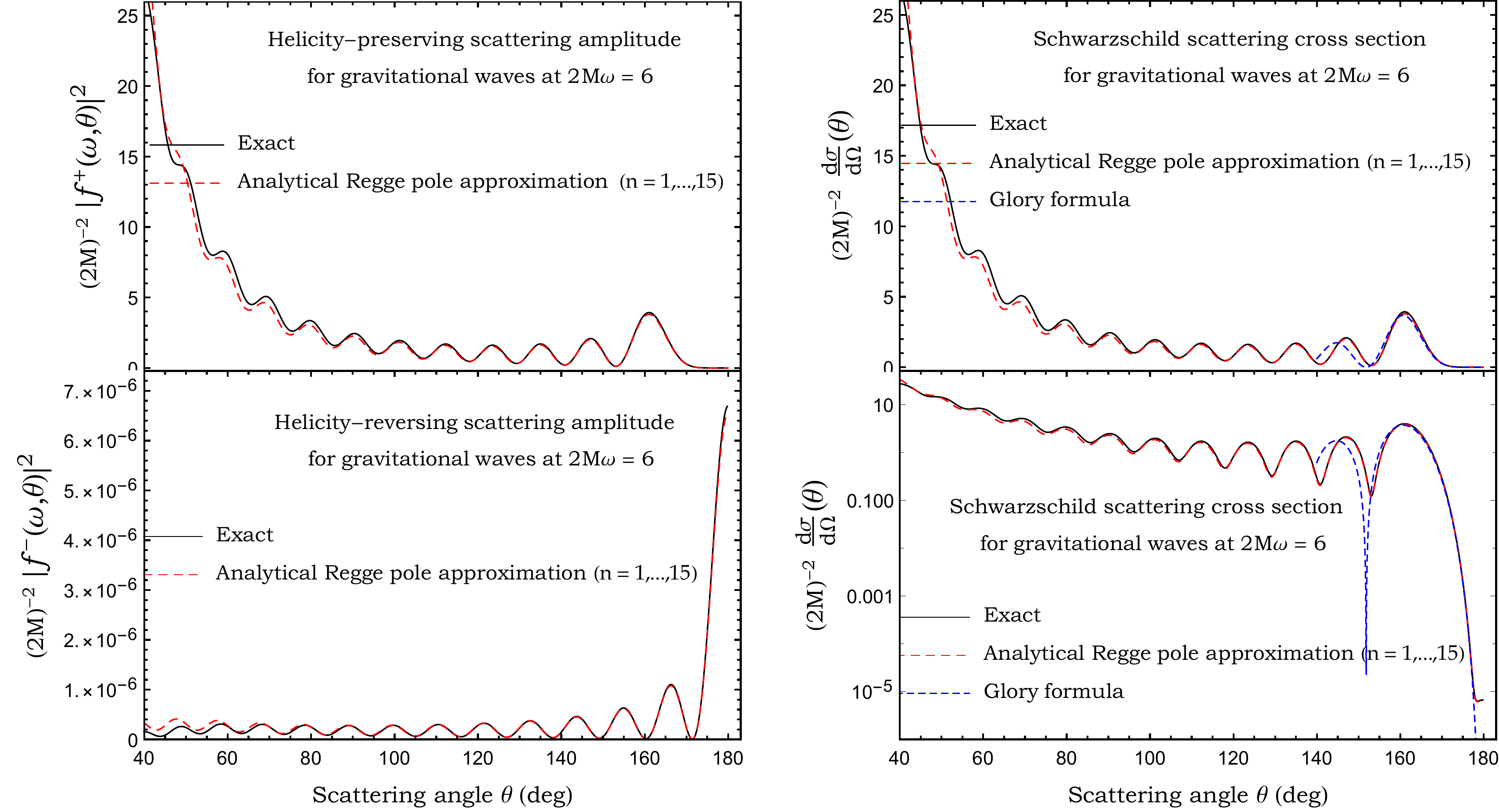}
\caption{\label{RP_approx_2Mw_6_s_2_n15} Scattering amplitudes and scattering cross section of a Schwarzschild BH for gravitational waves ($2M\omega=6$). We compare the exact results given in Sec.~\ref{SecIIa} with those obtained from the analytical Regge pole approximations constructed in Sec.~\ref{SecIV} where we sum over a large number of terms despite the inaccuracy of the approximations (\ref{PRapproxWKB}) and (\ref{Residues_approx}) for the higher Regge poles. We also display the glory cross section described by (\ref{Glory_scattering_formula}).}
\end{figure*}
\vfill

We now extend to the case of gravitational waves the approach developed in our previous paper, where we derived an analytical approximation fitting both the backward glory and a large part of the orbiting oscillations for the scattering of scalar and electromagnetic waves (see, Sec.~IV of Ref.~\cite{Folacci:2019cmc} for details and for our motivations). In order to obtain such an approximation for the helicity-preserving and helicity-reversing scattering amplitudes and for the total differential scattering cross section, we need asymptotic approximations at large reduced frequencies $2M\omega$ for the Regge poles $\lambda_n(\omega)$ and for the associated residues $r^{(e)}_n(\omega)$ and $r^{(o)}_n(\omega)$. We shall use the asymptotic form
\begin{equation}\label{PRapproxWKB}
\lambda_n(\omega) \approx 3\sqrt{3} \, M\omega + i \,(n-1/2) + \frac{\sqrt{3} \,
a_n}{18 M\omega}
\end{equation}
where
\begin{equation}\label{PRapproxWKB_denote1}
a_n =\frac{2}{3}\left[\frac{5}{12}(n-1/2)^2 + \frac{547}{144} \right]
\end{equation}
for the Regge poles \cite{Decanini:2009mu} and
\begin{eqnarray}\label{Residues_approx}
&& r_n^{(o)}(\omega) \approx  \frac{\left[-i \, 216 (3\sqrt{3} M\omega) / \xi \right]^{n-1/2}}{\sqrt{2\pi} \, (n-1)!} e^{i 2M\omega y} e^{i\pi \lambda_n(\omega)} \nonumber \\
& &
\end{eqnarray}
where
\begin{equation}\label{Residues_approx_xi_y}
\xi=7+4\sqrt{3} \quad \mathrm{and} \quad y=3-3\sqrt{3} + 4 \ln 2 - 3 \ln \xi
\end{equation}
for the odd residues \cite{DFOEH2019}. As far as the even residues are concerned, they can be obtained by inserting (\ref{Residues_approx}) into (\ref{resS_rel_CAM}). Here, it is important to recall that (\ref{PRapproxWKB}) and (\ref{Residues_approx}) which are accurate only for the lowest Regge poles and the associated residues are obtained by assuming the correspondence Regge poles/``surface waves'' propagating close to the photon sphere \cite{Andersson:1994rm,Decanini:2002ha,Decanini:2009mu,Dolan:2009nk,Decanini:2010fz}. From the technical point of view, (\ref{PRapproxWKB}) is derived by using
a WKB approximation to solve the Regge-Wheeler equation defined by Eqs.~(\ref{H_ZMetRW_equation}) and (\ref{pot Regge-Wheeler}) or, more precisely, by extending to Regge poles the approach developed in the context of the determination
of the QNMs by Schutz, Iyer and Will \cite{Schutz:1985zz,Iyer:1986vv,Iyer:1986nq} (see also Refs.~\cite{Dolan:2009nk,Decanini:2011eh}). Similarly, (\ref{Residues_approx}) has been derived in Ref.~\cite{DFOEH2019} by extending to Regge poles the calculations which have permitted to Dolan and Ottewill to derive an analytical expression for the QNM excitation factors \cite{Dolan:2011fh}.

By inserting now the approximations (\ref{PRapproxWKB}) for the Regge poles and (\ref{Residues_approx}) for the odd residues into the Regge pole sums (\ref{CAM_GW_Scattering_amp_decomp_RP}) where we take in addition into account (\ref{resS_rel_CAM}), we have at our disposal analytical approximations for the scattering amplitudes and for the cross section associated with gravitational waves which are formally valid for $2M\omega \to +\infty$. It should be however noted that the Regge pole sums can only involve a small number of terms because the approximations (\ref{PRapproxWKB}) and (\ref{Residues_approx}) are accurate only for the lowest Regge poles. As a consequence, from a theoretical point of view, our analytical Regge pole approximations cannot describe the cross sections for small scattering angles.

In Figs.~\ref{RP_approx_2Mw_3_s_2} and \ref{RP_approx_2Mw_6_s_2}, we compare the exact scattering amplitudes and the exact scattering cross section with their analytical approximations constructed below. The comparisons are achieved for the reduced frequencies $2M\omega=3$ and 6 and the summations are over the first five Regge poles. In these figures, we have in addition displayed the usual formula for the backward glory scattering cross section for gravitational waves which is given by \cite{Matzner:1985rjn}
\begin{equation}\label{Glory_scattering_formula}
  \left. \frac{d\sigma}{d\Omega}\right|_\mathrm{glory} = 30.752 M^3 \omega \, \left[J_4 \left(5.357 M\omega \sin \theta  \right)\right]^2,
\end{equation}
and which is formally valid for $2M\omega \gg 1$ and $|\theta -\pi| \ll 1$. Here, $J_4$ is a Bessel function of the first kind. It should be noted that (\ref{Glory_scattering_formula}) encodes only the contribution of the first backward glory and that the numerical factors appearing in this approximation are those obtained in Ref.~\cite{Crispino:2009ki}. We can observe that the analytical Regge pole approximations permit us to reproduce with very good agreement both the glory cross section [more precisely than (\ref{Glory_scattering_formula})] and a large part of the orbiting cross sections. Glory scattering and orbiting scattering are usually considered as two different effects \cite{Ford1959259,Handler:1980un,Matzner:1985rjn,Anninos:1992ih} and are described analytically by two different semiclassical analytic formulas (see Refs.~\cite{Matzner:1985rjn} and \cite{Anninos:1992ih} as well as Sec.~4.7.2 of Ref.~\cite{Frolov:1998wf} for a concise presentation). Here, we prove that it is possible from Regge pole sums to describe analytically both phenomena in a unique formula. In fact, by using the Regge pole approach of BH physics, we consider that glory and orbiting effects are not fundamentally different insofar as they are both generated by the excitation of surface waves propagating close to the BH photon sphere and are a consequence of diffractive effects due to this hypersurface \cite{Andersson:1994rm,Decanini:2002ha,Decanini:2009mu,Dolan:2009nk,Decanini:2010fz,DFOEH2019}. The asymptotic expressions (\ref{PRapproxWKB}) and (\ref{Residues_approx}) which are a direct consequence of this point of view and which are valid in the short-wavelength regime have permitted us to obtain this unified result.

In Fig.~\ref{RP_approx_2Mw_6_s_2_n15}, we now display the scattering amplitudes and the exact scattering cross section constructed from analytical Regge pole sums involving a large number of terms. Despite the inaccuracy of the approximations (\ref{PRapproxWKB}) and (\ref{Residues_approx}) for the higher Regge poles, it is surprising to observe that we are able to describe the scattering amplitudes and the cross section in a wide range of scattering angles.

\section{Conclusion and perspectives}
\label{Conc}

Scattering from BHs is usually is tackled from partial wave methods. We are developing an alternative approach based on the analytic extension of the $S$-matrix in the CAM plane and the use of Regge poles. It is very efficient in the short-wavelength regime where it allows us (i) to extract by resummation the information encoded in partial wave expansions and to overcome the difficulties linked to their lack of convergence due to the long-range nature of the fields propagating on BHs, (ii) to describe numerically, with an impressive agreement, the BH glory occurring in the backward direction as well as the orbiting oscillations appearing on the differential scattering cross sections for small and intermediate scattering angles and (iii) to describe semiclassically the BH glory and the orbiting oscillations by providing an accurate approximation that unifies these two phenomena (without the need for additional fitting parameters \cite{Anninos:1992ih}) and which is far superior to existing formulas \cite{Matzner:1985rjn,Anninos:1992ih}.

The CAM approach of scattering of massless fields by the Schwarzschild BH we have developed in our previous work \cite{Folacci:2019cmc} and in the present article is more general than it seems. Indeed, a major part of the formalism can be repeated identically in the context of scattering by four-dimensional, asymptotically flat, static spherically symmetric BHs (regular or not) as well as for some models of compact bodies described by Einstein's general relativity. In fact, if we consider a problem for which (i) the $S$-matrix has a behavior at infinity in the CAM plane which is identical to that of the Schwarzschild BH and (ii) the Regge pole spectrum has a structure in the CAM plane quite similar to that of the Schwarzschild BH, then only the numerical aspects of our work and, of course, the physical interpretation of the results must be modified or adapted.

We hope in next works to explore implications of our results in the context of strong gravitational lensing of electromagnetic and gravitational waves by BHs \cite{DFOEH2019}. A more challenging task is the extension of our study to scattering of waves by a Kerr BH. We also hope to make progress in this direction in the near future.

\begin{acknowledgments}

M.O.E.H. wishes to thank Sam Dolan for his kind invitation to the University of Sheffield where this work was completed. The authors are grateful to Sam Dolan for conversations and comments concerning this work and to Yves Decanini for discussions concerning the excitation of Regge modes.

\end{acknowledgments}

%\appendix*

\bibliography{BH_scattering_and_RP_GWs}

\begin{thebibliography}{36}%
\makeatletter
\providecommand \@ifxundefined [1]{%
 \@ifx{#1\undefined}
}%
\providecommand \@ifnum [1]{%
 \ifnum #1\expandafter \@firstoftwo
 \else \expandafter \@secondoftwo
 \fi
}%
\providecommand \@ifx [1]{%
 \ifx #1\expandafter \@firstoftwo
 \else \expandafter \@secondoftwo
 \fi
}%
\providecommand \natexlab [1]{#1}%
\providecommand \enquote  [1]{``#1''}%
\providecommand \bibnamefont  [1]{#1}%
\providecommand \bibfnamefont [1]{#1}%
\providecommand \citenamefont [1]{#1}%
\providecommand \href@noop [0]{\@secondoftwo}%
\providecommand \href [0]{\begingroup \@sanitize@url \@href}%
\providecommand \@href[1]{\@@startlink{#1}\@@href}%
\providecommand \@@href[1]{\endgroup#1\@@endlink}%
\providecommand \@sanitize@url [0]{\catcode `\\12\catcode `\$12\catcode
  `\&12\catcode `\#12\catcode `\^12\catcode `\_12\catcode `\%12\relax}%
\providecommand \@@startlink[1]{}%
\providecommand \@@endlink[0]{}%
\providecommand \url  [0]{\begingroup\@sanitize@url \@url }%
\providecommand \@url [1]{\endgroup\@href {#1}{\urlprefix }}%
\providecommand \urlprefix  [0]{URL }%
\providecommand \Eprint [0]{\href }%
\providecommand \doibase [0]{http://dx.doi.org/}%
\providecommand \selectlanguage [0]{\@gobble}%
\providecommand \bibinfo  [0]{\@secondoftwo}%
\providecommand \bibfield  [0]{\@secondoftwo}%
\providecommand \translation [1]{[#1]}%
\providecommand \BibitemOpen [0]{}%
\providecommand \bibitemStop [0]{}%
\providecommand \bibitemNoStop [0]{.\EOS\space}%
\providecommand \EOS [0]{\spacefactor3000\relax}%
\providecommand \BibitemShut  [1]{\csname bibitem#1\endcsname}%
\let\auto@bib@innerbib\@empty
%</preamble>
\bibitem [{\citenamefont {Folacci}\ and\ \citenamefont {Ould
  El~Hadj}(2019)}]{Folacci:2019cmc}%
  \BibitemOpen
  \bibfield  {author} {\bibinfo {author} {\bibfnamefont {A.}~\bibnamefont
  {Folacci}}\ and\ \bibinfo {author} {\bibfnamefont {M.}~\bibnamefont {Ould
  El~Hadj}},\ }\bibfield  {title} {\enquote {\bibinfo {title} {{Regge pole
  description of scattering of scalar and electromagnetic waves by a
  Schwarzschild black hole}},}\ }\href {\doibase 10.1103/PhysRevD.99.104079}
  {\bibfield  {journal} {\bibinfo  {journal} {Phys.\ Rev.\ D}\ }\textbf
  {\bibinfo {volume} {99}},\ \bibinfo {pages} {104079} (\bibinfo {year}
  {2019})},\ \Eprint {http://arxiv.org/abs/1901.03965} {arXiv:1901.03965
  [gr-qc]} \BibitemShut {NoStop}%
%%CITATION = ARXIV:1901.03965;%%
\bibitem [{\citenamefont {Matzner}\ and\ \citenamefont
  {Ryan}(1977)}]{Matzner:1977dn}%
  \BibitemOpen
  \bibfield  {author} {\bibinfo {author} {\bibfnamefont {R.~A.}\ \bibnamefont
  {Matzner}}\ and\ \bibinfo {author} {\bibfnamefont {M.~P.}\ \bibnamefont
  {Ryan}},\ }\bibfield  {title} {\enquote {\bibinfo {title} {{Low frequency
  limit of gravitational scattering}},}\ }\href {\doibase
  10.1103/PhysRevD.16.1636} {\bibfield  {journal} {\bibinfo  {journal} {Phys.\
  Rev.\ D}\ }\textbf {\bibinfo {volume} {16}},\ \bibinfo {pages} {1636}
  (\bibinfo {year} {1977})}\BibitemShut {NoStop}%
%%CITATION = PHRVA,D16,1636;%%
\bibitem [{\citenamefont {Matzner}\ and\ \citenamefont
  {Ryan}(1978)}]{MatznerRyan1978}%
  \BibitemOpen
  \bibfield  {author} {\bibinfo {author} {\bibfnamefont {R.~A.}\ \bibnamefont
  {Matzner}}\ and\ \bibinfo {author} {\bibfnamefont {M.~P.}\ \bibnamefont
  {Ryan}},\ }\bibfield  {title} {\enquote {\bibinfo {title} {{Scattering of
  gravitational radiation from vacuum black holes}},}\ }\href@noop {}
  {\bibfield  {journal} {\bibinfo  {journal} {Astrophys.\ J.\ Suppl.}\ }\textbf
  {\bibinfo {volume} {36}},\ \bibinfo {pages} {451} (\bibinfo {year}
  {1978})}\BibitemShut {NoStop}%
\bibitem [{\citenamefont {Handler}\ and\ \citenamefont
  {Matzner}(1980)}]{Handler:1980un}%
  \BibitemOpen
  \bibfield  {author} {\bibinfo {author} {\bibfnamefont {F.~A.}\ \bibnamefont
  {Handler}}\ and\ \bibinfo {author} {\bibfnamefont {R.~A.}\ \bibnamefont
  {Matzner}},\ }\bibfield  {title} {\enquote {\bibinfo {title} {{Gravitational
  wave scattering}},}\ }\href {\doibase 10.1103/PhysRevD.22.2331} {\bibfield
  {journal} {\bibinfo  {journal} {Phys.\ Rev.\ D}\ }\textbf {\bibinfo {volume}
  {22}},\ \bibinfo {pages} {2331} (\bibinfo {year} {1980})}\BibitemShut
  {NoStop}%
%%CITATION = PHRVA,D22,2331;%%
\bibitem [{\citenamefont {Dolan}(2008{\natexlab{a}})}]{Dolan:2007ut}%
  \BibitemOpen
  \bibfield  {author} {\bibinfo {author} {\bibfnamefont {S.~R.}\ \bibnamefont
  {Dolan}},\ }\bibfield  {title} {\enquote {\bibinfo {title} {{Scattering of
  long-wavelength gravitational waves}},}\ }\href {\doibase
  10.1103/PhysRevD.77.044004} {\bibfield  {journal} {\bibinfo  {journal}
  {Phys.\ Rev.\ D}\ }\textbf {\bibinfo {volume} {77}},\ \bibinfo {pages}
  {044004} (\bibinfo {year} {2008}{\natexlab{a}})},\ \Eprint
  {http://arxiv.org/abs/0710.4252} {arXiv:0710.4252 [gr-qc]} \BibitemShut
  {NoStop}%
%%CITATION = ARXIV:0710.4252;%%
\bibitem [{\citenamefont {Dolan}(2008{\natexlab{b}})}]{Dolan:2008kf}%
  \BibitemOpen
  \bibfield  {author} {\bibinfo {author} {\bibfnamefont {S.~R.}\ \bibnamefont
  {Dolan}},\ }\bibfield  {title} {\enquote {\bibinfo {title} {{Scattering and
  absorption of gravitational plane waves by rotating black holes}},}\ }\href
  {\doibase 10.1088/0264-9381/25/23/235002} {\bibfield  {journal} {\bibinfo
  {journal} {Class.\ Quant.\ Grav.}\ }\textbf {\bibinfo {volume} {25}},\
  \bibinfo {pages} {235002} (\bibinfo {year} {2008}{\natexlab{b}})},\ \Eprint
  {http://arxiv.org/abs/0801.3805} {arXiv:0801.3805 [gr-qc]} \BibitemShut
  {NoStop}%
%%CITATION = ARXIV:0801.3805;%%
\bibitem [{\citenamefont {Futterman}\ \emph {et~al.}(2012)\citenamefont
  {Futterman}, \citenamefont {Handler},\ and\ \citenamefont
  {Matzner}}]{Futterman:1988ni}%
  \BibitemOpen
  \bibfield  {author} {\bibinfo {author} {\bibfnamefont {J.~A.~H.}\
  \bibnamefont {Futterman}}, \bibinfo {author} {\bibfnamefont {F.~A.}\
  \bibnamefont {Handler}}, \ and\ \bibinfo {author} {\bibfnamefont {R.~A.}\
  \bibnamefont {Matzner}},\ }\href {\doibase 10.1017/CBO9780511735615} {\emph
  {\bibinfo {title} {{Scattering from Black Holes}}}},\ Cambridge Monographs on
  Mathematical Physics\ (\bibinfo  {publisher} {Cambridge University Press,
  Cambridge, England},\ \bibinfo {year} {2012})\BibitemShut {NoStop}%
%%CITATION = INSPIRE-266127;%%
\bibitem [{\citenamefont {Abbott}\ \emph {et~al.}(2016)\citenamefont {Abbott}
  \emph {et~al.}}]{Abbott:2016blz}%
  \BibitemOpen
  \bibfield  {author} {\bibinfo {author} {\bibfnamefont {B.~P.}\ \bibnamefont
  {Abbott}} \emph {et~al.} (\bibinfo {collaboration} {Virgo, LIGO
  Scientific}),\ }\bibfield  {title} {\enquote {\bibinfo {title} {{Observation
  of gravitational waves from a binary black hole merger}},}\ }\href {\doibase
  10.1103/PhysRevLett.116.061102} {\bibfield  {journal} {\bibinfo  {journal}
  {Phys.\ Rev.\ Lett.}\ }\textbf {\bibinfo {volume} {116}},\ \bibinfo {pages}
  {061102} (\bibinfo {year} {2016})},\ \Eprint
  {http://arxiv.org/abs/1602.03837} {arXiv:1602.03837 [gr-qc]} \BibitemShut
  {NoStop}%
%%CITATION = ARXIV:1602.03837;%%
\bibitem [{\citenamefont {Andersson}\ and\ \citenamefont
  {Thylwe}(1994)}]{Andersson:1994rk}%
  \BibitemOpen
  \bibfield  {author} {\bibinfo {author} {\bibfnamefont {N.}~\bibnamefont
  {Andersson}}\ and\ \bibinfo {author} {\bibfnamefont {K.~E.}\ \bibnamefont
  {Thylwe}},\ }\bibfield  {title} {\enquote {\bibinfo {title} {{Complex angular
  momentum approach to black hole scattering}},}\ }\href {\doibase
  10.1088/0264-9381/11/12/013} {\bibfield  {journal} {\bibinfo  {journal}
  {Class.\ Quant.\ Grav.}\ }\textbf {\bibinfo {volume} {11}},\ \bibinfo {pages}
  {2991} (\bibinfo {year} {1994})}\BibitemShut {NoStop}%
%%CITATION = CQGRD,11,2991;%%
\bibitem [{\citenamefont {Andersson}(1994)}]{Andersson:1994rm}%
  \BibitemOpen
  \bibfield  {author} {\bibinfo {author} {\bibfnamefont {N.}~\bibnamefont
  {Andersson}},\ }\bibfield  {title} {\enquote {\bibinfo {title} {{Complex
  angular momenta and the black hole glory}},}\ }\href {\doibase
  10.1088/0264-9381/11/12/014} {\bibfield  {journal} {\bibinfo  {journal}
  {Class.\ Quant.\ Grav.}\ }\textbf {\bibinfo {volume} {11}},\ \bibinfo {pages}
  {3003} (\bibinfo {year} {1994})}\BibitemShut {NoStop}%
%%CITATION = CQGRD,11,3003;%%
\bibitem [{\citenamefont {Decanini}\ \emph {et~al.}(2003)\citenamefont
  {Decanini}, \citenamefont {Folacci},\ and\ \citenamefont
  {Jensen}}]{Decanini:2002ha}%
  \BibitemOpen
  \bibfield  {author} {\bibinfo {author} {\bibfnamefont {Y.}~\bibnamefont
  {Decanini}}, \bibinfo {author} {\bibfnamefont {A.}~\bibnamefont {Folacci}}, \
  and\ \bibinfo {author} {\bibfnamefont {B.}~\bibnamefont {Jensen}},\
  }\bibfield  {title} {\enquote {\bibinfo {title} {{Complex angular momentum in
  black hole physics and the quasinormal modes}},}\ }\href {\doibase
  10.1103/PhysRevD.67.124017} {\bibfield  {journal} {\bibinfo  {journal}
  {Phys.\ Rev.\ D}\ }\textbf {\bibinfo {volume} {67}},\ \bibinfo {pages}
  {124017} (\bibinfo {year} {2003})},\ \Eprint
  {http://arxiv.org/abs/gr-qc/0212093} {arXiv:gr-qc/0212093} \BibitemShut
  {NoStop}%
\bibitem [{\citenamefont {Decanini}\ \emph {et~al.}(2010)\citenamefont
  {Decanini}, \citenamefont {Folacci},\ and\ \citenamefont
  {Raffaelli}}]{Decanini:2010fz}%
  \BibitemOpen
  \bibfield  {author} {\bibinfo {author} {\bibfnamefont {Y.}~\bibnamefont
  {Decanini}}, \bibinfo {author} {\bibfnamefont {A.}~\bibnamefont {Folacci}}, \
  and\ \bibinfo {author} {\bibfnamefont {B.}~\bibnamefont {Raffaelli}},\
  }\bibfield  {title} {\enquote {\bibinfo {title} {{Unstable circular null
  geodesics of static spherically symmetric black holes, Regge poles and
  quasinormal frequencies}},}\ }\href {\doibase 10.1103/PhysRevD.81.104039}
  {\bibfield  {journal} {\bibinfo  {journal} {Phys.\ Rev.\ D}\ }\textbf
  {\bibinfo {volume} {81}},\ \bibinfo {pages} {104039} (\bibinfo {year}
  {2010})},\ \Eprint {http://arxiv.org/abs/1002.0121} {arXiv:1002.0121 [gr-qc]}
  \BibitemShut {NoStop}%
%%CITATION = ARXIV:1002.0121;%%
\bibitem [{\citenamefont {Decanini}\ \emph
  {et~al.}(2011{\natexlab{a}})\citenamefont {Decanini}, \citenamefont
  {Esposito-Farese},\ and\ \citenamefont {Folacci}}]{Decanini:2011xi}%
  \BibitemOpen
  \bibfield  {author} {\bibinfo {author} {\bibfnamefont {Y.}~\bibnamefont
  {Decanini}}, \bibinfo {author} {\bibfnamefont {G.}~\bibnamefont
  {Esposito-Farese}}, \ and\ \bibinfo {author} {\bibfnamefont {A.}~\bibnamefont
  {Folacci}},\ }\bibfield  {title} {\enquote {\bibinfo {title} {{Universality
  of high-energy absorption cross sections for black holes}},}\ }\href
  {\doibase 10.1103/PhysRevD.83.044032} {\bibfield  {journal} {\bibinfo
  {journal} {Phys.\ Rev.\ D}\ }\textbf {\bibinfo {volume} {83}},\ \bibinfo
  {pages} {044032} (\bibinfo {year} {2011}{\natexlab{a}})},\ \Eprint
  {http://arxiv.org/abs/1101.0781} {arXiv:1101.0781 [gr-qc]} \BibitemShut
  {NoStop}%
%%CITATION = ARXIV:1101.0781;%%
\bibitem [{\citenamefont {Folacci}\ and\ \citenamefont {Ould
  El~Hadj}(2018)}]{Folacci:2018sef}%
  \BibitemOpen
  \bibfield  {author} {\bibinfo {author} {\bibfnamefont {A.}~\bibnamefont
  {Folacci}}\ and\ \bibinfo {author} {\bibfnamefont {M.}~\bibnamefont {Ould
  El~Hadj}},\ }\bibfield  {title} {\enquote {\bibinfo {title} {{Alternative
  description of gravitational radiation from black holes based on the Regge
  poles of the ${\cal S}$-matrix and the associated residues}},}\ }\href
  {\doibase 10.1103/PhysRevD.98.064052} {\bibfield  {journal} {\bibinfo
  {journal} {Phys.\ Rev.\ D}\ }\textbf {\bibinfo {volume} {98}},\ \bibinfo
  {pages} {064052} (\bibinfo {year} {2018})},\ \Eprint
  {http://arxiv.org/abs/1807.09056} {arXiv:1807.09056 [gr-qc]} \BibitemShut
  {NoStop}%
%%CITATION = ARXIV:1807.09056;%%
\bibitem [{\citenamefont {Watson}(1918)}]{Watson18}%
  \BibitemOpen
  \bibfield  {author} {\bibinfo {author} {\bibfnamefont {G.~N.}\ \bibnamefont
  {Watson}},\ }\bibfield  {title} {\enquote {\bibinfo {title} {{The diffraction
  of electric waves by the Earth}},}\ }\href@noop {} {\bibfield  {journal}
  {\bibinfo  {journal} {Proc.\ R.\ Soc.\ A}\ }\textbf {\bibinfo {volume}
  {95}},\ \bibinfo {pages} {83} (\bibinfo {year} {1918})}\BibitemShut {NoStop}%
\bibitem [{\citenamefont {Sommerfeld}(1949)}]{Sommerfeld49}%
  \BibitemOpen
  \bibfield  {author} {\bibinfo {author} {\bibfnamefont {A.}~\bibnamefont
  {Sommerfeld}},\ }\href@noop {} {\emph {\bibinfo {title} {Partial Differential
  Equations of Physics}}}\ (\bibinfo  {publisher} {Academic Press, New York},\
  \bibinfo {year} {1949})\BibitemShut {NoStop}%
\bibitem [{\citenamefont {Newton}(1982)}]{Newton:1982qc}%
  \BibitemOpen
  \bibfield  {author} {\bibinfo {author} {\bibfnamefont {R.~G.}\ \bibnamefont
  {Newton}},\ }\href@noop {} {\emph {\bibinfo {title} {{Scattering Theory of
  Waves and Particles}}}},\ \bibinfo {edition} {2nd}\ ed.\ (\bibinfo
  {publisher} {Springer-Verlag, New York},\ \bibinfo {year} {1982})\BibitemShut
  {NoStop}%
%%CITATION = INSPIRE-186478;%%
\bibitem [{\citenamefont {Decanini}\ and\ \citenamefont
  {Folacci}(2010)}]{Decanini:2009mu}%
  \BibitemOpen
  \bibfield  {author} {\bibinfo {author} {\bibfnamefont {Y.}~\bibnamefont
  {Decanini}}\ and\ \bibinfo {author} {\bibfnamefont {A.}~\bibnamefont
  {Folacci}},\ }\bibfield  {title} {\enquote {\bibinfo {title} {{Regge poles of
  the Schwarzschild black hole: A WKB approach}},}\ }\href {\doibase
  10.1103/PhysRevD.81.024031} {\bibfield  {journal} {\bibinfo  {journal}
  {Phys.\ Rev.\ D}\ }\textbf {\bibinfo {volume} {81}},\ \bibinfo {pages}
  {024031} (\bibinfo {year} {2010})},\ \Eprint {http://arxiv.org/abs/0906.2601}
  {arXiv:0906.2601 [gr-qc]} \BibitemShut {NoStop}%
%%CITATION = ARXIV:0906.2601;%%
\bibitem [{\citenamefont {Dolan}\ and\ \citenamefont
  {Ottewill}(2009)}]{Dolan:2009nk}%
  \BibitemOpen
  \bibfield  {author} {\bibinfo {author} {\bibfnamefont {S.~R.}\ \bibnamefont
  {Dolan}}\ and\ \bibinfo {author} {\bibfnamefont {A.~C.}\ \bibnamefont
  {Ottewill}},\ }\bibfield  {title} {\enquote {\bibinfo {title} {{On an
  expansion method for black hole quasinormal modes and Regge poles}},}\ }\href
  {\doibase 10.1088/0264-9381/26/22/225003} {\bibfield  {journal} {\bibinfo
  {journal} {Class.\ Quant.\ Grav.}\ }\textbf {\bibinfo {volume} {26}},\
  \bibinfo {pages} {225003} (\bibinfo {year} {2009})},\ \Eprint
  {http://arxiv.org/abs/0908.0329} {arXiv:0908.0329 [gr-qc]} \BibitemShut
  {NoStop}%
%%CITATION = ARXIV:0908.0329;%%
\bibitem [{\citenamefont {Regge}\ and\ \citenamefont
  {Wheeler}(1957)}]{Regge:1957td}%
  \BibitemOpen
  \bibfield  {author} {\bibinfo {author} {\bibfnamefont {T.}~\bibnamefont
  {Regge}}\ and\ \bibinfo {author} {\bibfnamefont {J.~A.}\ \bibnamefont
  {Wheeler}},\ }\bibfield  {title} {\enquote {\bibinfo {title} {{Stability of a
  Schwarzschild singularity}},}\ }\href {\doibase 10.1103/PhysRev.108.1063}
  {\bibfield  {journal} {\bibinfo  {journal} {Phys.\ Rev.}\ }\textbf {\bibinfo
  {volume} {108}},\ \bibinfo {pages} {1063--1069} (\bibinfo {year}
  {1957})}\BibitemShut {NoStop}%
%%CITATION = PHRVA,108,1063;%%
\bibitem [{\citenamefont {Zerilli}(1970)}]{Zerilli:1971wd}%
  \BibitemOpen
  \bibfield  {author} {\bibinfo {author} {\bibfnamefont {F.~J.}\ \bibnamefont
  {Zerilli}},\ }\bibfield  {title} {\enquote {\bibinfo {title} {{Gravitational
  field of a particle falling in a Schwarzschild geometry analyzed in tensor
  harmonics}},}\ }\href {\doibase 10.1103/PhysRevD.2.2141} {\bibfield
  {journal} {\bibinfo  {journal} {Phys.\ Rev.\ D}\ }\textbf {\bibinfo {volume}
  {2}},\ \bibinfo {pages} {2141--2160} (\bibinfo {year} {1970})}\BibitemShut
  {NoStop}%
%%CITATION = PHRVA,D2,2141;%%
\bibitem [{\citenamefont {Chandrasekhar}(1983)}]{Chandrasekhar:1985kt}%
  \BibitemOpen
  \bibfield  {author} {\bibinfo {author} {\bibfnamefont {S.}~\bibnamefont
  {Chandrasekhar}},\ }\href@noop {} {\emph {\bibinfo {title} {{The Mathematical
  Theory of Black Holes}}}}\ (\bibinfo  {publisher} {Oxford University Press,
  Oxford},\ \bibinfo {year} {1983})\BibitemShut {NoStop}%
%%CITATION = INSPIRE-224457;%%
\bibitem [{\citenamefont {Abramowitz}\ and\ \citenamefont
  {Stegun}(1965)}]{AS65}%
  \BibitemOpen
  \bibfield  {author} {\bibinfo {author} {\bibfnamefont {M.}~\bibnamefont
  {Abramowitz}}\ and\ \bibinfo {author} {\bibfnamefont {I.~A.}\ \bibnamefont
  {Stegun}},\ }\href@noop {} {\emph {\bibinfo {title} {Handbook of Mathematical
  Functions}}}\ (\bibinfo  {publisher} {Dover, New York},\ \bibinfo {year}
  {1965})\BibitemShut {NoStop}%
\bibitem [{\citenamefont {Chandrasekhar}\ and\ \citenamefont
  {Detweiler}(1975)}]{Chandrasekhar:1975zza}%
  \BibitemOpen
  \bibfield  {author} {\bibinfo {author} {\bibfnamefont {S.}~\bibnamefont
  {Chandrasekhar}}\ and\ \bibinfo {author} {\bibfnamefont {S.~L.}\ \bibnamefont
  {Detweiler}},\ }\bibfield  {title} {\enquote {\bibinfo {title} {{The
  quasi-normal modes of the Schwarzschild black hole}},}\ }\href {\doibase
  10.1098/rspa.1975.0112} {\bibfield  {journal} {\bibinfo  {journal} {Proc.\
  R.\ Soc.\ Lond.\ A}\ }\textbf {\bibinfo {volume} {344}},\ \bibinfo {pages}
  {441--452} (\bibinfo {year} {1975})}\BibitemShut {NoStop}%
%%CITATION = PRSLA,A344,441;%%
\bibitem [{\citenamefont {Inc.}()}]{Mathematica}%
  \BibitemOpen
  \bibfield  {author} {\bibinfo {author} {\bibfnamefont {Wolfram~Research{,}}\
  \bibnamefont {Inc.}},\ }\href@noop {} {\enquote {\bibinfo {title} {{\it
  Mathematica}, {V}ersion 10.0},}\ }\bibinfo {note} {(Wolfram Research{,} Inc.,
  Champaign, IL, 2014)}\BibitemShut {NoStop}%
\bibitem [{\citenamefont {Decanini}\ \emph {et~al.}()\citenamefont {Decanini},
  \citenamefont {Folacci},\ and\ \citenamefont {Ould El~Hadj}}]{DFOEH2019}%
  \BibitemOpen
  \bibfield  {author} {\bibinfo {author} {\bibfnamefont {Y.}~\bibnamefont
  {Decanini}}, \bibinfo {author} {\bibfnamefont {A.}~\bibnamefont {Folacci}}, \
  and\ \bibinfo {author} {\bibfnamefont {M.}~\bibnamefont {Ould El~Hadj}},\
  }\bibfield  {title} {\enquote {\bibinfo {title} {{Regge mode excitation and
  strong gravitational lensing (work in progress)}},}\ }\href@noop {} {\
  }\BibitemShut {NoStop}%
\bibitem [{\citenamefont {Schutz}\ and\ \citenamefont
  {Will}(1985)}]{Schutz:1985zz}%
  \BibitemOpen
  \bibfield  {author} {\bibinfo {author} {\bibfnamefont {B.~F.}\ \bibnamefont
  {Schutz}}\ and\ \bibinfo {author} {\bibfnamefont {C.~M.}\ \bibnamefont
  {Will}},\ }\bibfield  {title} {\enquote {\bibinfo {title} {{Black-hole normal
  modes: A semianalytic approach}},}\ }\href {\doibase 10.1086/184453}
  {\bibfield  {journal} {\bibinfo  {journal} {Astrophys.\ J.}\ }\textbf
  {\bibinfo {volume} {291}},\ \bibinfo {pages} {L33} (\bibinfo {year}
  {1985})}\BibitemShut {NoStop}%
%%CITATION = ASJOA,291,L33;%%
\bibitem [{\citenamefont {Iyer}\ and\ \citenamefont
  {Will}(1987)}]{Iyer:1986vv}%
  \BibitemOpen
  \bibfield  {author} {\bibinfo {author} {\bibfnamefont {S.}~\bibnamefont
  {Iyer}}\ and\ \bibinfo {author} {\bibfnamefont {C.~M.}\ \bibnamefont
  {Will}},\ }\bibfield  {title} {\enquote {\bibinfo {title} {{Black-hole normal
  modes: A WKB approach. I. Foundations and application of a higher-order WKB
  analysis of potential-barrier scattering}},}\ }\href {\doibase
  10.1103/PhysRevD.35.3621} {\bibfield  {journal} {\bibinfo  {journal} {Phys.\
  Rev.\ D}\ }\textbf {\bibinfo {volume} {35}},\ \bibinfo {pages} {3621}
  (\bibinfo {year} {1987})}\BibitemShut {NoStop}%
%%CITATION = PRINT-86-0935 (WASH.U.,ST.LOUIS);%%
\bibitem [{\citenamefont {Iyer}(1987)}]{Iyer:1986nq}%
  \BibitemOpen
  \bibfield  {author} {\bibinfo {author} {\bibfnamefont {S.}~\bibnamefont
  {Iyer}},\ }\bibfield  {title} {\enquote {\bibinfo {title} {{Black-hole normal
  modes: A WKB approach. II. Schwarzschild black holes}},}\ }\href {\doibase
  10.1103/PhysRevD.35.3632} {\bibfield  {journal} {\bibinfo  {journal} {Phys.\
  Rev.\ D}\ }\textbf {\bibinfo {volume} {35}},\ \bibinfo {pages} {3632}
  (\bibinfo {year} {1987})}\BibitemShut {NoStop}%
%%CITATION = PHRVA,D35,3632;%%
\bibitem [{\citenamefont {Decanini}\ \emph
  {et~al.}(2011{\natexlab{b}})\citenamefont {Decanini}, \citenamefont
  {Folacci},\ and\ \citenamefont {Raffaelli}}]{Decanini:2011eh}%
  \BibitemOpen
  \bibfield  {author} {\bibinfo {author} {\bibfnamefont {Y.}~\bibnamefont
  {Decanini}}, \bibinfo {author} {\bibfnamefont {A.}~\bibnamefont {Folacci}}, \
  and\ \bibinfo {author} {\bibfnamefont {B.}~\bibnamefont {Raffaelli}},\
  }\bibfield  {title} {\enquote {\bibinfo {title} {{Resonance and absorption
  spectra of the Schwarzschild black hole for massive scalar perturbations: a
  complex angular momentum analysis}},}\ }\href {\doibase
  10.1103/PhysRevD.84.084035} {\bibfield  {journal} {\bibinfo  {journal}
  {Phys.\ Rev.\ D}\ }\textbf {\bibinfo {volume} {84}},\ \bibinfo {pages}
  {084035} (\bibinfo {year} {2011}{\natexlab{b}})},\ \Eprint
  {http://arxiv.org/abs/1108.5076} {arXiv:1108.5076 [gr-qc]} \BibitemShut
  {NoStop}%
%%CITATION = ARXIV:1108.5076;%%
\bibitem [{\citenamefont {Dolan}\ and\ \citenamefont
  {Ottewill}(2011)}]{Dolan:2011fh}%
  \BibitemOpen
  \bibfield  {author} {\bibinfo {author} {\bibfnamefont {S.~R.}\ \bibnamefont
  {Dolan}}\ and\ \bibinfo {author} {\bibfnamefont {A.~C.}\ \bibnamefont
  {Ottewill}},\ }\bibfield  {title} {\enquote {\bibinfo {title} {{Wave
  Propagation and Quasinormal Mode Excitation on Schwarzschild Spacetime}},}\
  }\href {\doibase 10.1103/PhysRevD.84.104002} {\bibfield  {journal} {\bibinfo
  {journal} {Phys.\ Rev.\ D}\ }\textbf {\bibinfo {volume} {84}},\ \bibinfo
  {pages} {104002} (\bibinfo {year} {2011})},\ \Eprint
  {http://arxiv.org/abs/1106.4318} {arXiv:1106.4318 [gr-qc]} \BibitemShut
  {NoStop}%
%%CITATION = ARXIV:1106.4318;%%
\bibitem [{\citenamefont {Matzner}\ \emph {et~al.}(1985)\citenamefont
  {Matzner}, \citenamefont {DeWitt-Morette}, \citenamefont {Nelson},\ and\
  \citenamefont {Zhang}}]{Matzner:1985rjn}%
  \BibitemOpen
  \bibfield  {author} {\bibinfo {author} {\bibfnamefont {R.~A.}\ \bibnamefont
  {Matzner}}, \bibinfo {author} {\bibfnamefont {C.}~\bibnamefont
  {DeWitt-Morette}}, \bibinfo {author} {\bibfnamefont {B.}~\bibnamefont
  {Nelson}}, \ and\ \bibinfo {author} {\bibfnamefont {T.-R.}\ \bibnamefont
  {Zhang}},\ }\bibfield  {title} {\enquote {\bibinfo {title} {{Glory scattering
  by black holes}},}\ }\href {\doibase 10.1103/PhysRevD.31.1869} {\bibfield
  {journal} {\bibinfo  {journal} {Phys.\ Rev.\ D}\ }\textbf {\bibinfo {volume}
  {31}},\ \bibinfo {pages} {1869} (\bibinfo {year} {1985})}\BibitemShut
  {NoStop}%
%%CITATION = PHRVA,D31,1869;%%
\bibitem [{\citenamefont {Crispino}\ \emph {et~al.}(2009)\citenamefont
  {Crispino}, \citenamefont {Dolan},\ and\ \citenamefont
  {Oliveira}}]{Crispino:2009ki}%
  \BibitemOpen
  \bibfield  {author} {\bibinfo {author} {\bibfnamefont {L.~C.~B.}\
  \bibnamefont {Crispino}}, \bibinfo {author} {\bibfnamefont {S.~R.}\
  \bibnamefont {Dolan}}, \ and\ \bibinfo {author} {\bibfnamefont {E.~S.}\
  \bibnamefont {Oliveira}},\ }\bibfield  {title} {\enquote {\bibinfo {title}
  {{Scattering of massless scalar waves by Reissner-Nordstr\"om black
  holes}},}\ }\href {\doibase 10.1103/PhysRevD.79.064022} {\bibfield  {journal}
  {\bibinfo  {journal} {Phys.\ Rev.\ D}\ }\textbf {\bibinfo {volume} {79}},\
  \bibinfo {pages} {064022} (\bibinfo {year} {2009})},\ \Eprint
  {http://arxiv.org/abs/0904.0999} {arXiv:0904.0999 [gr-qc]} \BibitemShut
  {NoStop}%
%%CITATION = ARXIV:0904.0999;%%
\bibitem [{\citenamefont {Ford}\ and\ \citenamefont
  {Wheeler}(1959)}]{Ford1959259}%
  \BibitemOpen
  \bibfield  {author} {\bibinfo {author} {\bibfnamefont {K.~W.}\ \bibnamefont
  {Ford}}\ and\ \bibinfo {author} {\bibfnamefont {J.~A.}\ \bibnamefont
  {Wheeler}},\ }\bibfield  {title} {\enquote {\bibinfo {title} {Semiclassical
  description of scattering},}\ }\href {\doibase 10.1016/0003-4916(59)90026-0}
  {\bibfield  {journal} {\bibinfo  {journal} {Annals of Physics}\ }\textbf
  {\bibinfo {volume} {7}},\ \bibinfo {pages} {259} (\bibinfo {year}
  {1959})}\BibitemShut {NoStop}%
\bibitem [{\citenamefont {Anninos}\ \emph {et~al.}(1992)\citenamefont
  {Anninos}, \citenamefont {DeWitt-Morette}, \citenamefont {Matzner},
  \citenamefont {Yioutas},\ and\ \citenamefont {Zhang}}]{Anninos:1992ih}%
  \BibitemOpen
  \bibfield  {author} {\bibinfo {author} {\bibfnamefont {P.}~\bibnamefont
  {Anninos}}, \bibinfo {author} {\bibfnamefont {C.}~\bibnamefont
  {DeWitt-Morette}}, \bibinfo {author} {\bibfnamefont {R.~A.}\ \bibnamefont
  {Matzner}}, \bibinfo {author} {\bibfnamefont {P.}~\bibnamefont {Yioutas}}, \
  and\ \bibinfo {author} {\bibfnamefont {T.~R.}\ \bibnamefont {Zhang}},\
  }\bibfield  {title} {\enquote {\bibinfo {title} {{Orbiting cross-sections:
  Application to black hole scattering}},}\ }\href {\doibase
  10.1103/PhysRevD.46.4477} {\bibfield  {journal} {\bibinfo  {journal} {Phys.\
  Rev.\ D}\ }\textbf {\bibinfo {volume} {46}},\ \bibinfo {pages} {4477}
  (\bibinfo {year} {1992})}\BibitemShut {NoStop}%
%%CITATION = PHRVA,D46,4477;%%
\bibitem [{\citenamefont {Frolov}\ and\ \citenamefont
  {Novikov}(1998)}]{Frolov:1998wf}%
  \BibitemOpen
  \bibfield  {author} {\bibinfo {author} {\bibfnamefont {V.~P.}\ \bibnamefont
  {Frolov}}\ and\ \bibinfo {author} {\bibfnamefont {I.~D.}\ \bibnamefont
  {Novikov}},\ }\href {\doibase 10.1007/978-94-011-5139-9} {\emph {\bibinfo
  {title} {{Black Hole Physics: Basic Concepts and New Developments}}}},\
  \bibinfo {series} {Fundamental Theories of Physics}, Vol.~\bibinfo {volume}
  {96}\ (\bibinfo  {publisher} {Kluwer Academic Publishers, Dordrecht, The
  Netherlands},\ \bibinfo {year} {1998})\BibitemShut {NoStop}%
%%CITATION = FTPHD,96,;%%
\end{thebibliography}%

\end{document}